\documentclass[a4paper,11pt]{article}
\pdfoutput=1 

\usepackage{bbold}
\usepackage{amsmath}
\usepackage{amssymb}
\usepackage{dsfont}
\usepackage{graphicx}
\usepackage{xspace}
\usepackage[multiple]{footmisc}
\usepackage{bbm} 
\usepackage{bm}
\usepackage{color}
\usepackage[utf8]{inputenc}
\usepackage{cancel}
\usepackage{slashed}
\usepackage{soul}
\usepackage{units}
\usepackage{enumitem}
\usepackage{makecell}
\usepackage{nicefrac}
\usepackage{mathtools}
\usepackage{subfig}
\usepackage{float}
\usepackage{cite}
\usepackage[normalem]{ulem}
\usepackage{tikz} 
\usetikzlibrary{chains,shapes,fit,calc}
\usetikzlibrary{positioning,arrows}

\usetikzlibrary{patterns,decorations.pathreplacing}
\usetikzlibrary{decorations.pathmorphing}
\usetikzlibrary{decorations.markings}
\usetikzlibrary{arrows,shapes}
\usetikzlibrary{shapes.misc}
\tikzset{
    gluon/.style={decorate, draw=black,
        decoration={coil,amplitude=4pt, segment length=4pt,aspect=0.7}} 
}
\tikzset{
    photon/.style={decorate, decoration={snake}},
}

\usepackage[normalem]{ulem}

\newcommand{\DRb}{\overline{\mbox{DR}}}

\newcommand{\beq}{\begin{eqnarray}}
\newcommand{\eeq}{\end{eqnarray}}
\newcommand{\ba}{\begin{eqnarray}}
\newcommand{\ea}{\end{eqnarray}}
\newcommand{\be}{\begin{equation}}
\newcommand{\ee}{\end{equation}}
\newcommand{\bpmatrix}{\begin{pmatrix}}
\newcommand{\epmatrix}{\end{pmatrix}}

\newcommand\NMSSMCALC{{\tt NMSSMCALC}\xspace}
\newcommand\SARAH{{\tt SARAH}\xspace}

\newcommand\TARCER{{\tt TARCER}\xspace}

\newcommand\FeynArts{{\tt FeynArts}\xspace}
\newcommand\FeynCalc{{\tt FeynCalc}\xspace}

\newcommand{\order}[1]{\mathcal{O}(#1)}

\newcommand{\vev}{\textit{v}}

\newcommand{\vs}{\textit{v}_s}
\newcommand{\MSbar}{\overline{\text{MS}}}
\newcommand{\DRbar}{\overline{\text{DR}}}

\newcommand{\del}{\partial}

\newcommand{\doublet}[2]{\begin{pmatrix} #1 \\ #2 \end{pmatrix}}

\newcommand{\ccdot}{\!\cdot\!}

\renewcommand{\Re}{\text{Re}}
\renewcommand{\Im}{\text{Im}}

\newcommand{\s}{\newline \vspace*{-3.5mm}}

\newcommand{\tb}{t_{\beta}}

\newcommand{\mueff}{\mu_{\text{eff}}}

\newcommand{\bea}{\begin{eqnarray}}
\newcommand{\eea}{\end{eqnarray}}

\renewcommand{\eqref}[1]{Eq.~(\ref{#1})}

\RequirePackage[colorlinks=false
  ,linktocpage=true
  ,pdfproducer=medialab
  ,pdfa=true
]{hyperref}
\usepackage[capitalise, english]{cleveref}

\crefname{section}{Sec.}{Secs.}
\crefname{table}{Tab.}{Tabs.}

\graphicspath{{figs/}}
\makeatletter
\def\input@path{{chapters/}}
\makeatother

\begin{document}
\vspace{1cm}

\title{
\vspace*{-3cm}
\phantom{h} \hfill\mbox{\small DESY-22-141}\\[-1.1cm]
\phantom{h} \hfill\mbox{\small KA-TP-23-2022}\\[1cm]
The Trilinear Higgs Self-Couplings at 
  $\order{\alpha_t^2}$ \\ in the CP-Violating NMSSM}

\newcommand{\AddrKAITP}{
Institute for Theoretical Physics (ITP), 
Karlsruhe Institute of Technology, \\
Wolfgang-Gaede-Stra{\ss}e 1, D-76131 Karlsruhe, Germany}

\author{
Christoph Borschensky$^{1\,}$\footnote{E-mail:
  \texttt{christoph.borschensky@kit.edu}}, 
Thi Nhung Dao$^{2\,}$\footnote{E-mail:
  \texttt{nhung.daothi@phenikaa-uni.edu.vn}},
Martin
Gabelmann$^{3\,}$\footnote{E-mail: \texttt{martin.gabelmann@desy.de}}, \\
Margarete
M\"uhlleitner$^{1\,}$\footnote{E-mail: \texttt{margarete.muehlleitner@kit.edu}},
Heidi Rzehak$^{4\,}$\footnote{E-mail: \texttt{heidi.rzehak@itp.uni-tuebingen.de}}
\\[9mm]
{\small\it 
$^1$Institute for Theoretical Physics, Karlsruhe Institute of Technology,} \\
{\small\it Wolfgang-Gaede-Str. 1, 76131 Karlsruhe, Germany}\\[3mm]
{\small\it
$^2$Faculty of Fundamental Sciences, PHENIKAA University, Hanoi 12116,
Vietnam}\\[3mm]
{\small\it 
$^3$Deutsches Elektronen-Synchrotron DESY, Notkestr.~85,}\\
{\small\it 22607 Hamburg, Germany}\\[3mm]
{\small \it 
$^4$Institute for Theoretical Physics, University of T\"ubingen,}\\{\small \it Auf der Morgenstelle 14, 72076 T\"ubingen, Germany}\\[3mm]
}
\maketitle

\begin{abstract}
In supersymmetric theories the Higgs boson masses are derived
quantities where higher-order corrections have to be included in order
to match the measured Higgs mass value at the precision of current 
experiments. Closely related through the 
Higgs potential are the Higgs self-interactions. In addition, the measurement of
the trilinear Higgs self-coupling provides the first step towards the
reconstruction of the Higgs potential and the experimental
verification of the Higgs mechanism {\it sui generis}. In this paper, we advance our
prediction of the trilinear Higgs self-couplings in the CP-violating
Next-to-Minimal Supersymmetric extension of the SM (NMSSM). We provide the
${\cal O}(\alpha_t^2)$ corrections in the gaugeless limit at vanishing
external momenta. The higher-order corrections turn out to be larger
than the corresponding mass corrections but show the expected
perturbative convergence. The inclusion of the loop-corrected
effective trilinear Higgs self-coupling in gluon fusion into Higgs
pairs and the estimate of the theoretical uncertainty due to missing
higher-order corrections indicate that the missing electroweak higher-order
corrections may be significant.
\end{abstract}

\thispagestyle{empty}
\vfill
\pagebreak

\tableofcontents

\section{Introduction}
\label{sec:intro}
The measurement of the trilinear Higgs self-coupling is one of the
most important tasks at the LHC and future colliders
\cite{DiMicco:2019ngk}. It is the first step towards the experimental
reconstruction of the Higgs potential and hence the direct experimental
verification of the Higgs mechanism {\it sui generis}
\cite{Djouadi:1999gv,Djouadi:1999rca,Muhlleitner:2000jj}. At the LHC, it is accessible in
gluon fusion into Higgs pairs. In models beyond the Standard Model (SM) with extended
Higgs sectors, the Higgs self-couplings are also involved in
Higgs-to-Higgs decays. Through the Higgs potential, the
trilinear Higgs self-coupling is related to the Higgs boson
mass. While in the SM the Higgs mass is an ad hoc input
parameter, in supersymmetric theories \cite{Golfand:1971iw,
  Volkov:1973ix, Wess:1974tw, Fayet:1974pd,Fayet:1977yc, Fayet:1976cr,
  Nilles:1982dy,Nilles:1983ge,
  Frere:1983ag,Derendinger:1983bz,Haber:1984rc,
  Sohnius:1985qm,Gunion:1984yn, Gunion:1986nh} it is derived from the
parameters of the model. In the Minimal Supersymmetric extension of the
SM (MSSM)
\cite{Gunion:1989we,Martin:1997ns,Dawson:1997tz,Djouadi:2005gj} the
Higgs quartic couplings are given
in terms of the gauge couplings leading to an upper bound of the
tree-level mass of the order of the
$Z$ boson mass so that considerable higher-order corrections are
required to shift the SM-like Higgs boson mass to
the observed value of 125.09~GeV \cite{HiggsAtlas}. 
In the Next-to-MSSM (NMSSM)
\cite{Barbieri:1982eh,Dine:1981rt,Ellis:1988er,Drees:1988fc,Ellwanger:1993xa,Ellwanger:1995ru,Ellwanger:1996gw,Elliott:1994ht,King:1995vk,Franke:1995tc,Maniatis:2009re,Ellwanger:2009dp}
the situation is somewhat more relaxed due to the tree-level
contribution stemming from the inclusion of the additional complex
singlet field. In the last years, a lot of effort has been put to
provide precise predictions for the Higgs mass values at higher loop
level both in the MSSM and the NMSSM. For recent reviews, see
\cite{Slavich:2020zjv,R:2021bml}. For the trilinear Higgs self-couplings the
corresponding corrections have not yet been provided at the same level
of precision as for the masses. In the MSSM, the one-loop corrections to the effective
trilinear couplings have been provided many years ago in
\cite{Barger:1991ed,Hollik:2001px,Dobado:2002jz}. The
process-dependent corrections to heavy scalar MSSM Higgs decays into a
lighter Higgs pair have been calculated in
\cite{Williams:2007dc,Williams:2011bu}. The two-loop ${\cal
  O}(\alpha_t \alpha_s)$ SUSY-QCD
corrections to the top/stop-loop induced corrections have been made
available within the effective potential approach in
\cite{Brucherseifer:2013qva}. In the NMSSM, we provided the full
one-loop corrections for the CP-conserving NMSSM \cite{Nhung:2013lpa}. They are
sizeable so that the inclusion of the two-loop corrections is
mandatory to reduce the theoretical uncertainties due to missing
higher-order corrections. Consequently, we subsequently calculated the two-loop
${\cal O}(\alpha_t \alpha_s)$ corrections in the limit of vanishing
external momenta in \cite{Muhlleitner:2015dua}, in the CP-violating
NMSSM. The full one-loop corrections to the Higgs-to-Higgs decays and
other on-shell two-body decays were implemented 
 in \cite{Baglio:2019nlc}. For
corrections to the trilinear Higgs self-couplings in
non-supersymmetric (non-SUSY) Higgs models, see for example Refs.~\cite{Kanemura:2002vm,Kanemura:2004mg,Kanemura:2015mxa,Kanemura:2017wtm,Basler:2017uxn,Basler:2019iuu} for
one-loop and
Refs.~\cite{Senaha:2018xek,Braathen:2019pxr,Braathen:2019zoh,Bahl:2022jnx}
for two-loop results, and
Refs.~\cite{Krause:2016oke,Bojarski:2015kra,Krause:2017mal,Krause:2018wmo,Denner:2018opp,Krause:2019oar,Krause:2019qwe,Azevedo:2021ylf,Egle:2022wmq,Goodsell:2017pdq}
for the process-dependent Higgs-to-Higgs decays at one-loop level. \s

In this paper we continue our effort in increasing the precision
for the predictions of the trilinear Higgs self-couplings in the context of
the NMSSM. We calculate the two-loop corrections at
${\cal O}(\alpha_t^2)$ to the trilinear Higgs self-couplings of the
complex NMSSM. They are obtained in the limit of zero external momenta
and vanishing gauge couplings. We consistently apply the same
renormalisation schemes as in our computation of the loop-corrections
to the Higgs boson masses, based on a mixed on-shell-$\DRbar$
renormalisation in the Higgs sector and the possibility to choose
between on-shell (OS) and $\DRbar$ renormalisation in the top/stop
sector. Our corrections have been implemented in our Fortran code {\tt
  NMSSMCALC} \cite{Baglio:2013iia,King:2015oxa} and can be downloaded from the URL:
\begin{center}
\url{https://www.itp.kit.edu/~maggie/NMSSMCALC/}
\end{center}

The paper is organized as follows. In Sec.~\ref{sec:model}
we introduce the tree-level sectors of the NMSSM that are
relevant for our calculation and set up our notation. In
Sec.~\ref{sec:coupling} we give the definitions for the
loop-corrected effective trilinear Higgs self-couplings and for the loop
corrections to the Higgs-to-Higgs decays. We specify the
approximations that we apply and the renormalisation schemes that we
use. In Sec.~\ref{sec:pheno} we briefly present the set-up of our
numerical analysis and the scan that we
performed. Sections~\ref{sec:bp2os} to \ref{sec:higgspair} are
dedicated to our numerical analysis. In Sec.~\ref{sec:bp2os} we
discuss the impact of our corrections on the effective trilinear Higgs
self-couplings and on the Higgs-to-Higgs decays for two specific
parameter points, in Sec.~\ref{sec:scatter} we investigate these
effects for our whole sample to get a more general picture. The
effects of our corrections in the context of Higgs pair production are
analysed in Sec.~\ref{sec:higgspair}. Our conclusions are given in Sec.~\ref{sec:concl}.

\section{The Tree-Level NMSSM}
\label{sec:model}
In order to set our notation we briefly introduce the two sectors
relevant for the renormalisation, the Higgs and the 
top/stop sectors. While the computation of the ${\cal O}(\alpha_t^2)$ contributions to
the Higgs self-energies and tadpoles involves neutralinos and charginos, they do not
need to be renormalised as vertices and propagators with these particles
only enter at the two-loop level. For the definition of the
electroweakino masses and mixing angles in the gaugeless limit we
refer to Ref.~\cite{Dao:2019qaz}. We work in the $\mathbb{Z}_3$
symmetric NMSSM including CP violation. For the computation of the
two-loop corrections to the trilinear Higgs self-couplings of the
neutral Higgs bosons at ${\cal O} (\alpha_t \alpha_s + \alpha_t^2)$ we
apply the gaugeless limit and follow the notation of our previous
calculations~\cite{Muhlleitner:2014vsa,Dao:2019qaz,Muhlleitner:2015dua,Dao:2021khm}. Note
that in contrast to the MSSM, the trilinear and quartic Higgs
couplings in the NMSSM do not vanish in the gaugeless limit, but
involve the parameters $\lambda,\kappa, A_\lambda,A_\kappa$. The NMSSM superpotential is given by 
\begin{align}
    \mathcal{W}_{\text{NMSSM}} = 
	&
		\left[y_e \hat{H}_d\ccdot \hat{L} \hat{E}^c 
		+ y_d  \hat{H}_d \ccdot \hat{Q} \hat{D}^c 
		- y_u \hat{H}_u \ccdot \hat{Q} \hat{U}^c
		\right]  - \lambda \hat{S} \hat{H}_d\ccdot \hat{H}_u 
		+ \frac{1}{3} \kappa \hat{S}^3  \;,
    \label{eq:wnmssm}
\end{align}
in terms of the quark and lepton superfields $\hat{Q}$, $\hat{U}$,
$\hat{D}$, $\hat{L}$, $\hat{E}$, the Higgs doublet
superfields $\hat{H}_d$, $\hat{H}_u$ and the singlet superfield
$\hat{S}$. Charge conjugated fields are denoted by the superscript
$c$. We have suppressed color and generation indices for better readability. The
symplectic product $x\ccdot y= \epsilon_{ij}x^iy^j$ ($i,j=1,2$) is built with the
anti-symmetric tensor $\epsilon_{12}=\epsilon^{12}=1$. 
Working in the CP-violating NMSSM, the parameters $\lambda,\kappa$ are
in general complex. All $y_x$ ($x=e,d,u$) are taken to
  be real by rephasing the left- and right-handed Weyl-spinor fields
  as $x_{L,R}\to x_{L,R} e^{i  \varphi_{\text{\tiny L,R}}}$. In
  our computation, the Yukawa couplings $y_x$ are assumed
  to be diagonal in flavour space, and we only include $y_t$ while all
  other Yukawa couplings are set to zero. 
  The soft SUSY breaking Lagrangian is given by
\begin{eqnarray}
{\cal L}_{\text{soft},\text{ NMSSM}} &=& -m_{H_d}^2 H_d^\dagger H_d - m_{H_u}^2
H_u^\dagger H_u -
m_{\tilde{Q}}^2 \tilde{Q}^\dagger \tilde{Q} - m_{\tilde{L}}^2 \tilde{L}^\dagger \tilde{L}
- m_{\tilde{u}_R}^2 \tilde{u}_R^* 
\tilde{u}_R - m_{\tilde{d}_R}^2 \tilde{d}_R^* \tilde{d}_R 
\nonumber \\\nonumber
&& - m_{\tilde{e}_R}^2 \tilde{e}_R^* \tilde{e}_R - ( [y_e A_e H_d\ccdot
\tilde{L} \tilde{e}_R^* + y_d
A_d H_d\ccdot \tilde{Q} \tilde{d}_R^* - y_u A_u H_u \ccdot \tilde{Q}
\tilde{u}_R^*] + \mathrm{h.c.}) \\
&& -\frac{1}{2}(M_1 \tilde{B}\tilde{B} + M_2
\tilde{W}_i\tilde{W}_i + M_3 \tilde{G}\tilde{G} + \mathrm{h.c.}) \nonumber
\\ 
\label{eq:lagrangiansoft}
&&- m_S^2 |S|^2 +
(\lambda 
A_\lambda S H_d  \ccdot  H_u - \frac{1}{3} \kappa
A_\kappa S^3 + \mathrm{h.c.}) \;,
\end{eqnarray}
where again quark and lepton generation indices are suppressed. The $\tilde{Q}$, {$\tilde{u}_R$, $\tilde{d}_R$ and $\tilde{L}$, $\tilde{e}_R$} denote the complex scalar
components of the corresponding quark and lepton superfields. 
The soft
SUSY breaking gaugino mass parameters $M_i$ ($i=1,2,3$) of the bino,
wino and gluino fields $\tilde{B}$, $\tilde{W}_i$ ($i=1,2,3$) and
$\tilde{G}$ as well as 
the soft SUSY breaking trilinear couplings
$A_x$ ($x=\lambda,\kappa,u,d,e$) are complex in the CP-violating
NMSSM in contrast to 
the soft SUSY breaking mass parameters of the scalar fields, 
$m_X^2$ ($X=S,H_d,H_u,\tilde{Q},\tilde{u}_R,\tilde{d}_R,\tilde{L},\tilde{e}_R$),
which are real. 

\subsection{The Higgs Boson Sector \label{sec:higgssector}}
The tree-level Higgs boson potential is obtained from 
${\mathcal{L}}_{\text{soft},\text{ NMSSM}}$, the $F$-terms of
$\mathcal{W}_{\text{NMSSM}}$ and  the $D$-terms originating
from the gauge sector,
\beq
V_{H}  &=& (|\lambda S|^2 + m_{H_d}^2)H_d^\dagger H_d+ (|\lambda S|^2
+ m_{H_u}^2)H_u^\dagger H_u +m_S^2 |S|^2 \nonumber \\
&& + \frac{1}{8} (g_2^2+g_1^{2})(H_d^\dagger H_d-H_u^\dagger H_u )^2
+\frac{1}{2} g_2^2|H_d^\dagger H_u|^2 \nonumber \\ 
&&   + |\kappa S^2- \lambda  H_{d} \ccdot H_{u}  |^2+
\big[\frac{1}{3} \kappa
A_{\kappa} S^3-\lambda A_\lambda S   H_{d}\ccdot  H_{u}  +\mathrm{h.c.} \big] \;.
  \label{eq:higgspotential}
\eeq
In the gaugeless limit, the $U(1)_Y$ and $SU(2)_L$ gauge
couplings $g_1\to 0$ and  $g_2\to 0$ while $\tan\theta_W =g_2/g_1$
is kept constant, with $\theta_W $ being the weak mixing angle. This
is equivalent to the  limit of vanishing electric charge and
tree-level vector boson masses, $e,M_W,M_Z\to 0$, while keeping
$\tan\theta_W$ constant. 
\s

The Higgs boson fields are expanded around their vacuum expectation
values (VEVs) $v_u$, $v_d$, and $v_s$ as
\begin{equation}
    H_d = \doublet{\frac{v_d + h_d +i a_d}{\sqrt 2}}{h_d^-}, \,\, 
    H_u = e^{i\varphi_u}\doublet{h_u^+}{\frac{v_u + h_u +i a_u}{\sqrt 2}},\,\,
    S= \frac{e^{i\varphi_s}}{\sqrt 2} (v_s + h_s + ia_s)\, ,
   \label{eq:vevs}
\end{equation}
where $\varphi_{u,s}$ denote the CP-violating phases. We can replace the three VEVs
by $\tan\beta$, the SM VEV $v$ and the effective $\mu$ parameter $\mueff$ as
\begin{align} 
\tb&\equiv\tan\beta= v_u/v_d \\
\vev^2&=v_u^2+v_d^2\approx \left(\unit[246]{GeV}\right)^2 \\
    \mueff&=\frac{e^{i \varphi_s}}{\sqrt 2}\vs \lambda\,.
    \label{eq:mueffcalc}
\end{align}
Note that the MSSM limit is smoothly retraced by taking the limit
$\lambda,\kappa\to 0,~v_s\to\infty$ and at the same time 
keeping $\mueff$ and  $\kappa/\lambda$ constant.
From the Higgs potential of Eq.~(\ref{eq:higgspotential}) we obtain 
the tree-level tadpoles, the Higgs mass
matrices and the trilinear Higgs self-couplings. For the tadpole
coefficients we have
\beq
(\bm{t})_l = t_{{\bm{\phi}_l}} = \frac{\del V_H}{\del {{\bm{\phi}_l}}}\bigg|_{\bm{\phi}=0}, \,\, l = 1, \dots, 6
  \,\, ,
\eeq
with
\beq
{\bm{\phi}}=(h_d,h_u,h_s,a_d,a_u,a_s)^T \;.
\eeq
Only five of the tadpoles are independent, since $t_{a_u}=t_{a_d}/t_\beta$.
The neutral Higgs mass matrix in the interaction basis is obtained
as
\beq
{\cal M}_{\phi_l \phi_m}= \frac{\del^2 V_H}{\del {\bm{\phi}_l} \del {\bm{\phi}_m}}\bigg|_{\bm{\phi}=0}
\eeq
and the charged one as ($r,s=1,2$)
\beq
{\cal M}_{h^+_r h^-_s}= \frac{\del V_H}{\del {\bm{h}^{c,\dagger}_r} \del
  {\bm{h}^c_s}}\bigg|_{\bm{h}^c=0} \;, \quad \mbox{with} \; \bm{h} ^c=  (h_d^{-*},h_u^+) \;.
\eeq
The trilinear couplings  which need to be renormalised at two-loop level for the
calculation of the ${\cal O}(\alpha_t^2)$ corrections in this paper, are
obtained as
\beq
\lambda_{\bm{\phi}_l \bm{\phi}_m \bm{\phi}_n} \equiv \lambda_{lmn} = 
\frac{\del^3 V_H}{\del {\bm{\phi}_l} \del {\bm{\phi}_m} \del
  {\bm{\phi}_n}}\bigg|_{\bm{\phi}=0} \;.
\eeq
The explicit expressions for the tadpoles and the squared mass matrices
$\mathcal{M}_{\phi \phi}$ and $\mathcal{M}_{h^+ h^-}$ are given in
Ref. \cite{Dao:2019qaz} and those for the trilinear Higgs
self-couplings can be found in the Appendix of
Ref.~\cite{Muhlleitner:2015dua}. 
The neutral Higgs mass eigenstates are obtained by a two-fold rotation
that first separates the Goldstone component through the rotation
$\mathcal{R}^G(\beta_n)$, {\it i.e.}~it transforms from the basis
$(h_d, h_u, h_s, a_d, a_u,a_s)$ to 
$(h_d, h_u, h_s, a, a_s, G^0)$,
and afterwards rotates into the mass basis
$(h_1, h_2, h_3, h_4, h_5, G^0)$
with the rotation matrix ${\cal R}$, 
\begin{align}
    \mathcal{M}_{hh}  = \,\, & \mathcal{R}^G(\beta_n)\mathcal{M}_{\phi\phi}(\mathcal{R}^G(\beta_n))^T  \label{eq:massmatrix} \\
  \mathcal{M}_{hh}^\prime  =\,\, & \mathcal{R} \mathcal{M}_{hh}
                                   \mathcal{R}^{T} 
  \label{eq:massmatrix2}\\
     =\,\, &
             \text{diag}(m_{h_1}^2,m_{h_2}^2,m_{h_3}^2,m_{h_4}^2,m_{h_5}^2,m_{G^0}^2)
             \,, \nonumber
\end{align}
where the neutral Goldstone boson mass is equal to the $Z$ boson mass,
$m_{G^0} =M_Z$, in 't Hooft Feynman gauge. It vanishes in the gaugeless limit.
The charged Higgs fields are rotated to the mass eigenstates with a
single rotation $\mathcal{R}^{G^-}(\beta_c)$,
\begin{align}
\mathcal{R}^{G^-}(\beta_c)\mathcal{M}_{h^+h^-}(\mathcal{R}^{G^-}(\beta_c))^T
=
\text{diag}(m_{G^\pm}^2, M_{H^\pm}^2)
    \, . \label{eq:chargedHiggsmass}
\end{align}
In the 't Hooft Feynman gauge the charged Goldstone boson mass is
equal to the charged $W$ boson mass, $m_{G^\pm} =M_W$, and vanishes in the 
gaugeless limit.
At tree-level the rotation angles $\beta_n$ and $\beta_c$ coincide with $\beta$, $\beta_c=\beta_n=\beta$. 
They are distinguished here
as $\beta_n$ and $\beta_c$ are mixing angles and do not need to obtain
a counterterm which is not the case for $\beta$ that arises from the 
ratio of the VEVs. It has to be renormalised and receives a
non-vanishing counterterm. 
After the renormalisation they are set equal to the tree-level value of $\beta$ again.
Our tree-level masses are denoted by small letters $m$, apart from the
charged Higgs boson mass. When we talk about loop-corrected masses,
they are denoted by capital $M$. In our renormalisation of the
trilinear coupling we will adapt the same renormalisation conditions as
those used in the two-loop corrections of the masses. 
\s

\newcommand{\ReAkappa}{\Re A_{\kappa}}
\newcommand{\ReAlambda}{\Re A_{\lambda}}
\newcommand{\cosks}{c_{\varphi_\omega}}
\newcommand{\sinks}{s_{\varphi_\omega}}

We apply the SUSY Les Houches Accord (SLHA)
\cite{Skands:2003cj,Allanach:2008qq} and in accordance with this
accord decompose the complex parameters $A_\lambda$
and $A_\kappa$ into their imaginary and real parts. While in our
program code {\tt NMSSMCALC} also $\lambda$ and $\kappa$ are read in
in terms of their real and complex part in accordance with the SLHA,
internally, we choose a different, more convenient,
parametrisation. We decompose $\lambda$ and $\kappa$ into their absolute values and phases
$\varphi_\lambda$ and $\varphi_\kappa$. We note that the  phases
enter the tree-level Higgs mass matrix 
in two combinations together with $\varphi_u$ and $\varphi_s$,
\begin{eqnarray}
\varphi_y &=& \varphi_\kappa - \varphi_\lambda + 2 \varphi_s -
\varphi_u \label{eq:phase1} \\
\varphi_w &=& \varphi_\kappa + 3 \varphi_s \;, \label{eq:phase2}
\end{eqnarray}
where $\varphi_y$ is the only CP-violating phase at tree level in
the Higgs sector. If $\varphi_y=0$,  the CP-even components, $h_u,h_d,
h_s$, hence do not mix with the CP-odd ones, $a_d,a_u,a_s$.  
We use the tadpole conditions to replace $\Im A_{\lambda,\kappa}$ as well 
as $m_{H_{u,d},S}^2$ by the tadpole parameters $t_{a_d, a_s}$ and
$t_{h_{d,u,s}}$, respectively, {\it cf.}~Ref.~\cite{Dao:2019qaz} for
details. \s

In \NMSSMCALC, we have two possibilities to choose the set of input parameters in the Higgs sector, either
\begin{align}
\left\{ t_{h_d},t_{h_u},t_{h_s},t_{a_d},t_{a_s},M_{H^\pm}^2,v,s_{\theta_W},
e,\tan\beta,|\lambda|,v_s,|\kappa|,\ReAkappa,\varphi_\lambda,\varphi_\kappa,\varphi_u,\varphi_s
\right\} \,, \label{eq:inputset1}
\end{align}
or
\begin{align}
\left\{ t_{h_d},t_{h_u},t_{h_s},t_{a_d},t_{a_s},v,s_{\theta_W},
e,\tan\beta,|\lambda|,v_s,|\kappa|,\ReAlambda,\ReAkappa,\varphi_\lambda,\varphi_\kappa,\varphi_u,\varphi_s
\right\} \,. \label{eq:inputset2}
\end{align}
In the first choice the charged Higgs mass is an input parameter while
in the second one we have $\ReAlambda$ as an input
parameter. 

\subsection{The Top/Stop Sector}
For the calculation of the Higgs self-couplings at the order ${\cal
O}(\alpha_t^2)$, the top/stop sector needs to be renormalised at
${\cal O}(\alpha_t)$. The top mass and the top-quark Yukawa coupling are related as, 
\begin{equation}
    m_t = \frac{v_u y_t}{\sqrt 2} 
          e^{i(\varphi_u+\varphi_{\text{\tiny L}}-\varphi_{\text{\tiny R}})}\,,
          \label{eq:mttree}
\end{equation}
with $m_t$ and $y_t$ being real in our convention. Applying the
freedom of choice of the phases $\varphi_{\text{\tiny L}}$,
$\varphi_{\text{\tiny R}}$ of the left- and right-handed 
top-quark fields, we define $\varphi_{\text{\tiny
    L}}=-\varphi_{\text{\tiny R}}= -\varphi_u/2$. Thereby the stop
mass matrix in the $(\tilde{t}_{L},\tilde{t}_{R})^T$ basis in the
gaugeless limit is given by
\begin{align}
 {\cal M}_{\tilde{t}}  &=
     \begin{pmatrix}
             m_{\tilde{Q}_3}^2 + m_t^2 & m_t
             \left(A_t^*e^{-i\varphi_u}-\frac{\mueff}{\tan\beta}\right) \\[2mm]
             m_t \left(A_t e^{i\varphi_u}
               -\frac{\mueff^*}{\tan\beta}\right) &
             m_{\tilde{t}_R}^2+m_t^2 
         \end{pmatrix} \\
    \text{diag}( m_{\tilde{t}_1}^2, m_{\tilde{t}_2}^2)& =
    \mathcal{U}^{\tilde t}  {\cal M}_{\tilde{t}}  {\mathcal{U}^{\tilde{t}}}^\dagger\,,
\end{align}
where $\mathcal{U}^{\tilde{t}}$ denotes the rotation matrix for the
left- and right-handed 
stop fields $\tilde{t}_{L,R}$ into the mass eigenstates
$\tilde{t}_{1,2}$. We set the bottom quark mass to zero everywhere so
that the right-handed sbottom states decouple and only left-handed
sbottom states appear in the computation. In the stop sector the
parameters to be renormalised at one-loop level are
\be 
m_t, \; m_{\tilde{Q}_3}, \; m_{\tilde t_R} \quad \mbox{and} \quad A_t \;.
\label{eq:stopparset} 
\ee

\section{The Loop-Corrected Couplings \label{sec:coupling}}

\subsection{Definition}
The renormalised trilinear Higgs self-coupling $\hat{\lambda}_{ijk}$
at two-loop order between the interaction states $h_i$, $h_j$ and $h_k$ is given by
\beq
\hat{\lambda}_{ijk} &=& \lambda_{ijk} + \Delta^{(1)} \lambda_{ijk} +
\Delta^{(2)} \lambda_{ijk} \;.
\eeq
Here the indices $i,j,k$ refer to the interaction basis $(h_d,h_u,h_s,a_d,a_u,a_s)$.
We denote the trilinear tree-level Higgs self-coupling by $\lambda_{ijk}$
and the one- and two-loop corrections to it 
by $\Delta^{(1)}\lambda_{ijk}$ and $\Delta^{(2)}\lambda_{ijk}$, respectively. The
explicit expressions for the tree-level couplings in the interaction
basis are given in App.~A of \cite{Muhlleitner:2015dua}. \s

Applying the description in \cite{Muhlleitner:2015dua}, we define the
so-called effective trilinear Higgs self-couplings as follows. Both
one-loop and two-loop corrections are computed in the approximation
of zero external momenta, more specifically: 
 \begin{itemize}
 \item  In the one-loop corrections, $\Delta^{(1)}\lambda$ (for
   simplicity, here and in the following we drop the indices '$ijk$'
   where they are not needed), we include
   only corrections at ${\cal O}(\alpha_t)$. These are coming from the
   top/stop sector and are hence the dominant ones. They have been
   discussed in detail in \cite{Muhlleitner:2015dua}.  
 \item For the two-loop corrections, $\Delta^{(2)}\lambda$, we include
   the dominant contributions of ${\cal O}(\alpha_t\alpha_s)$ and
   ${\cal O}(\alpha_t^2)$,  
 \beq 
\Delta^{(2)} \lambda= \Delta^{\alpha_t\alpha_s}
\lambda+\Delta^{\alpha_t^2} \lambda \;,
\eeq 
where the QCD corrections $ \Delta^{\alpha_t\alpha_s} \lambda$
have been computed in  \cite{Muhlleitner:2015dua}. In this paper the
$\Delta^{\alpha_t^2} \lambda$ corrections are calculated for the
first time. 
 \item After calculating $\hat{\lambda}_{ijk}$ in the interaction
   basis $(h_d,h_u,h_s,a_d,a_u,a_s)$ we rotate it first to the basis
   $(h_d,h_u,h_s,a,a_s,G^0)$ to single out the couplings with the neutral
   Goldstone bosons as 
\beq  
\hat{\hat{\lambda}}_{nmq}=
{\cal R}^G_{ni}{\cal R}^G_{mj} {\cal R}^G_{qk}
\hat{\lambda}_{ijk} \;.
\label{eq:singling}
\eeq 
\item To obtain the effective trilinear couplings in the mass
  eigenstate basis we use the loop-corrected rotation matrix ${\cal
    R}^{l,\text{eff}}$. This matrix diagonalizes the loop-corrected mass matrix evaluated 
in the approximation of vanishing external momentum,
\beq  
\hat{\lambda}^{\text{eff}}_{abc}= {\cal R}^{l,\text{eff}}_{an}\, {\cal
  R}^{l,\text{eff}}_{bm}\, {\cal R}^{l,\text{eff}}_{cq}\,
\hat{\hat{\lambda}}_{nmq} \;, 
\label{eq:effdef}
\eeq 
where the indices $a,b,c$ refer to the mass basis
$(H_1,H_2,H_3,H_4,H_5)$ and $n,m,q$ to the interaction basis
$(h_d,h_u,h_s,a,a_s)$. The matrix ${\cal R}^{l,\text{eff}}$ is a $5\times 5$ matrix
and returned as an SLHA output of
{\tt NMSSMCALC} in the block {\tt NMHMIXC}. Note that we denote the
loop-corrected Higgs boson masses by capital letters ($H_i$) and the
tree-level ones by lower letters ($h_i$).
 \end{itemize}

We will later also calculate the Higgs-to-Higgs decays where
we have to ensure the proper on-shell conditions of the external
Higgs bosons. In this case the one-loop corrections $\Delta^{(1)}\lambda$ include the 
full electroweak corrections together with non-vanishing momentum effects. 
They have been computed by us in the context of the CP-conserving and CP-violating
NMSSM in Ref.~\cite{Nhung:2013lpa} and Ref.~\cite{Muhlleitner:2015dua}, respectively. 
 The two-loop part $\Delta^{(2)} \lambda$ 
 contains the ${\cal O}(\alpha_t\alpha_s)$ and  ${\cal
   O}(\alpha_t^2)$ part, computed in the zero-momentum approximation as
 described above. Throughout, at two-loop order we
   apply the gaugeless limit. In order to ensure the proper on-shell
 conditions of the Higgs bosons, to the maximum extent
   possible in the context of our calculation, the amplitude for the decay
 process $H_a\to  H_b+H_c$ is computed by including the effect from
 the finite wave-function 
 renormalisation factor matrix $\bf{Z}$ which is defined by  
\be 
{\cal M}_{H_a\to H_b+H_c} = {\cal R}^l_{an} {\cal R}^l_{bm}{\cal
   R}^l_{cq} \hat{\hat{\lambda}}_{nmq} \;, \label{eq:decayamplitude}
\ee 
where 
\beq
{\cal R}^l= \bf{Z} {\cal R} \;,\label{eq:zwfrmatrix}
\eeq 
with ${\cal R}$ being the matrix that rotates 
the interaction eigenstates $(h_d,h_u,h_s,a,a_s,G^0)$ to the tree-level
mass eigenstates $(h_1,h_2,h_3,h_4,h_5,G^0)$. The
definition of the matrix $\bf{Z}$ can be found in Ref.~\cite{Nhung:2013lpa}
for the CP-conserving case and Ref.~\cite{Baglio:2019nlc} for the
CP-violating case.\footnote{For the complex MSSM this has been derived
in Ref.~\cite{Williams:2007dc}.} 

\subsection{One- and Two-Loop Corrections}
To be consistent, we compute the one- and two-loop corrections to the
trilinear Higgs self-couplings in accordance with the corresponding
one- and two-loop corrections to the Higgs boson masses. This means we
use the same 
renormalisation conditions in the
higher-order corrections to the trilinear couplings as the ones we
used in our computation for the masses. For our mass calculations, the
detailed presentation of the one-loop corrections can be found in
\cite{Ender:2011qh,Graf:2012hh} and of the 
two-loop corrections up to order ${\cal O}(\alpha_t \alpha_s)$
in \cite{Muhlleitner:2014vsa}, to order ${\cal O}(
\alpha_t^2)$ in \cite{Dao:2019qaz} and to order ${\cal O}
((\alpha_t+\alpha_\lambda+\alpha_\kappa)^2)$ in \cite{Dao:2021khm},
together with the corresponding renormalisation conditions and the
explicit definitions of the counterterms. The
one-loop corrections to the trilinear Higgs self-couplings in the real
NMSSM have been presented in \cite{Nhung:2013lpa} and to two-loop order ${\cal
  O}(\alpha_t \alpha_s)$ in the CP-violating NMSSM in
\cite{Muhlleitner:2015dua}. Throughout our computations we apply a
mixed on-shell (OS)-$\overline{\mbox{DR}}$ renormalisation scheme. In
the two-loop corrections which require the renormalisation of the
top/stop sector we provide the option to choose between OS and
$\overline{\mbox{DR}}$ renormalisation. All details can be found in
the respective papers. Here we focus on a minimal description
and refer the reader for further information to this literature. \s

In case the charged Higgs mass is used as independent input the
parameters related to the Higgs sector that need to
be renormalised are given by\footnote{Note, that for the two-loop
  ${\cal O}(\alpha_t^2)$ corrections computed in this publication, we only need to
renormalise the parameters of the top/stop sector.}  
\beq
\underbrace{ t_{h_d},t_{h_u},t_{h_s},t_{a_d},t_{a_s},M_{H^\pm}^2,v}_{\mbox{on-shell
 scheme}},
\underbrace{\tan\beta,|\lambda|,v_s,|\kappa|,\ReAkappa,\varphi_\lambda,\varphi_\kappa,\varphi_u,\varphi_s}_{\overline{\mbox{DR}} \mbox{ scheme}}\,,
\label{eq:mixedcond1}
\eeq
and by
\beq
\underbrace{ t_{h_d},t_{h_u},t_{h_s},t_{a_d},t_{a_s},v}_{\mbox{on-shell
 scheme}},
\underbrace{\tan\beta,|\lambda|,v_s,|\kappa|,\ReAlambda,\ReAkappa,\varphi_\lambda,\varphi_\kappa,\varphi_u,\varphi_s}_{\overline{\mbox{DR}} \mbox{ scheme}}\,,
\label{eq:mixedcond2}
\eeq
for $\mbox{Re}(A_\lambda)$ as independent input. Note, that if we
apply the gaugeless limit we do not need to renormalise the neutral and charged gauge
boson masses, $M_Z$ and $M_W$, and the electric coupling
$e$. For the Higgs fields, which need to be renormalised as well, we
choose $\overline{\mbox{DR}}$ conditions. The details of the
renormalisation procedure and the counterterms are given in the above
mentioned papers so that we do not repeat them here. \s

The one-loop corrections $\Delta^{(1)}\lambda$ of the trilinear Higgs
self-couplings can be decomposed as
\beq
\Delta^{(1)}\lambda = \Delta^{(1)}\lambda^{\text{UR}} +
\Delta^{(1)}\lambda^{\text{CT}} \;,
\eeq
where the first term denotes the unrenormalised part given by the genuine 
one-loop diagrams. For the ${\cal O}(\alpha_t)$ correction, they
comprise the one-loop diagrams with top and stops running in the
loops and we restrict ourselves to the gaugeless limit. For the trilinear
couplings used in the Higgs-to-Higgs decays we include the complete one-loop
corrections at non-vanishing gauge couplings. 
The explicit expressions for the order ${\cal O}(\alpha_t)$
corrections to the trilinear self-couplings are given in 
App.~B and the counterterm expressions
$\Delta^{(1)}\lambda^{\text{CT}}$ are given in App.~C of
Ref.~\cite{Muhlleitner:2015dua}. \s

\begin{figure}[t!]
    \centering
\includegraphics[width=0.23\textwidth]{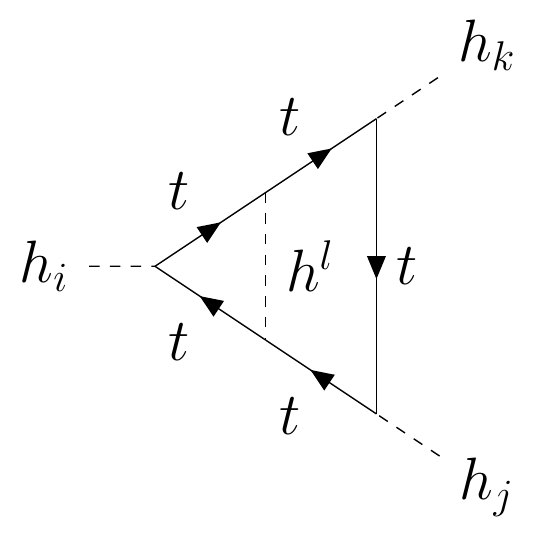}
\includegraphics[width=0.23\textwidth]{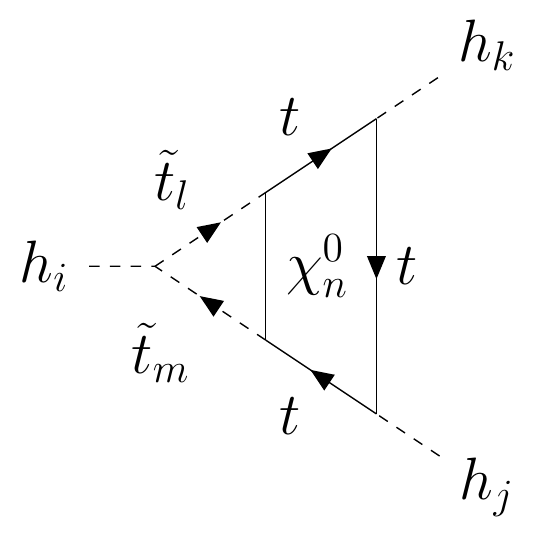} 
\includegraphics[width=0.23\textwidth]{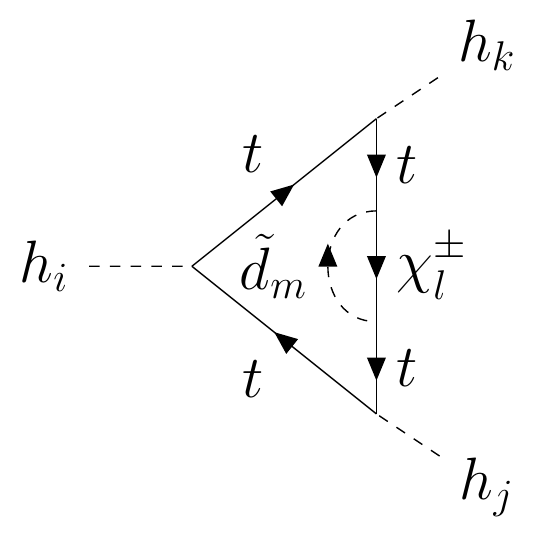}
\includegraphics[width=0.23\textwidth]{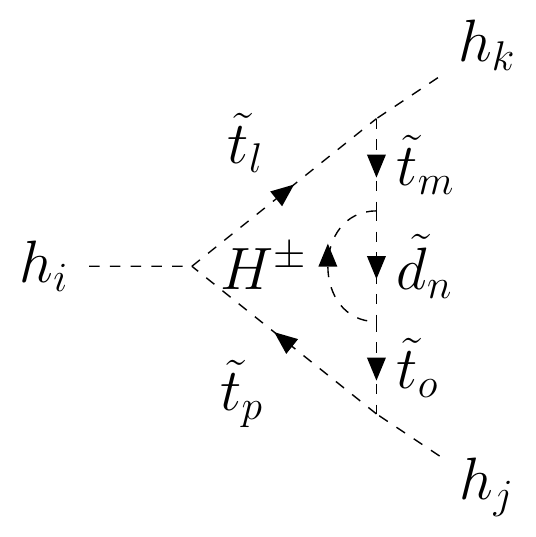} 
    \caption{Sample diagrams contributing to the trilinear Higgs
      self-coupling at ${\cal O}(\alpha_t^2)$. \label{fig:oalth2}}
\end{figure}

The two-loop corrections $\Delta^{(2)}\lambda$ of the trilinear Higgs
self-couplings are composed of 
\beq
\Delta^{(2)}\lambda = \Delta^{(2)}\lambda^{\text{UR}} +
\Delta^{(2)}\lambda^{\text{CT1L}} + \Delta^{(2)}\lambda^{\text{CT2L}} \;.
\eeq
The unrenormalised part $\Delta^{(2)}\lambda^{\text{UR}}$ consists of
the genuine two-loop diagrams contributing at order ${\cal O}(\alpha_t
\alpha_s)$ and ${\cal O}(\alpha_t^2)$. Some sample diagrams for the
newly computed ${\cal O}(\alpha_t^2)$ are depicted in
Fig.~\ref{fig:oalth2}. In the
approximation of zero external momenta all two-loop three-point
functions can be written in terms of products of one-loop integrals or
the two-loop tadpole integral. Their analytic
expressions are given in the literature
\cite{Davydychev:1992mt,Ford:1992pn,Scharf:1993ds,Weiglein:1993hd,Berends:1994ed,Martin:2001vx,Martin:2005qm}. The
counterterm contributions $\Delta^{(2)}\lambda^{\text{CT1L}}$ arise
from one-loop diagrams containing top and stop contributions combined 
with one insertion of a counterterm of the order ${\cal O} (\alpha_s)$
(for the ${\cal O}(\alpha_t \alpha_s)$ corrections)
or the order ${\cal O} (\alpha_t)$ (for the ${\cal O}(\alpha_t^2)$ corrections)
from the top/stop sector. The counterterm contribution
$\Delta^{(2)}\lambda^{\text{CT2L}}$ consists of the ${\cal O}(\alpha_t
\alpha_s)$ and ${\cal O}(\alpha_t^2)$ counterterms and
is manifestly zero when only top/stop contributions are
considered.

\section{Set-up of the Calculation and of the Numerical Analysis}
\label{sec:pheno}
\subsection{Tools, Checks and {\tt NMSSMCALC} Release\label{sec:tools}}
For the computation of the loop-corrected trilinear Higgs
self-couplings we made use of our setup for our 
computation of the loop-corrected Higgs masses
\cite{Dao:2021khm}. There we used \SARAH
4.14.3 \cite{Staub:2008uz,Staub:2010jh,Staub:2012pb,Staub:2013tta,Goodsell:2014bna,Goodsell:2014pla} to generate the model file including the vertex
coun\-ter\-terms. The file was then used in \FeynArts
3.10 \cite{Kublbeck:1990xc,Hahn:2000kx} to generate all required one- and
two-loop Feynman diagrams for the calculation of the corrections to
the trilinear Higgs self-couplings. The evaluation of the
fermion traces and the tensor reduction of the one- and two-loop integrals
and the amplitudes with the counterterm-inserted diagrams was done
with the help of \FeynCalc 9.2.0
\cite{Mertig:1990an,Shtabovenko:2016sxi} and its \TARCER
plugin~\cite{Mertig:1998vk}. We performed
three independent calculations which all agreed. We also explicitly
checked the ultraviolet (UV)-finiteness of the loop-corrected Higgs self-couplings. 
\s

The calculation of the trilinear Higgs self-couplings at one- and two-loop order
as well as the Higgs-to-Higgs decays including these corrections, has
been implemented in {\tt NMSSMCALC} \cite{Baglio:2013iia} both for the
CP-conserving and the CP-violating 
NMSSM. The new {\tt NMSSMCALC} version 5.1 can be
downloaded from the URL:  
\begin{center}
\url{https://www.itp.kit.edu/~maggie/NMSSMCALC/}
\end{center}
The input file {\tt inp.dat} includes the option to choose between the
different loop orders in the trilinear couplings and correspondingly
the Higgs-to-Higgs-decay widths. The effective trilinear Higgs
self-couplings as defined above are given out in the output
file. 

\subsection{The Parameter Scan}
For the numerical discussion of our results we used the data set that
we had generated for Ref.~\cite{Dao:2021khm} by performing a scan in
the NMSSM parameter space and keeping only those data sets that are in
accordance with the relevant experimental constraints. We briefly
summarise them here for convenience of the reader. We ensured 
compatibility with experimental constraints from the Higgs data by using {\tt
   HiggsBounds} 5.9.0~\cite{Bechtle:2008jh,Bechtle:2011sb,Bechtle:2013wla} 
and {\tt HiggsSignals}
2.6.1~\cite{Bechtle:2013xfa}. The required effective
NMSSM Higgs boson couplings normalised to the corresponding
SM values were generated with {\tt NMSSMCALC}. 
For valid points, $\chi^2$ computed by {\tt
  HiggsSignals}-2.6.1 needs to be consistent with  
an SM $\chi^2$ within $2\sigma$.\footnote{In {\tt HiggsSignals}-2.6.1,
  the SM $\chi^2$ obtained with the latest data set is 84.44. We
  allowed the NMSSM $\chi^2$ to be in the range $[78.26,90.62]$.}  
For this analysis, we checked the sample again with
  the updated {\tt HiggsBounds} 5.10.2 and {\tt HiggsSignals} 2.6.2\footnote{In {\tt HiggsSignals}-2.6.2, the SM $\chi^2$ obtained with the latest data set is 89.62.} 
  and found that more than 90\% of the points (and in particular the two
  benchmark points discussed below) are still in the allowed region.
We required one of the neutral CP-even Higgs bosons, called $h$ from now on,
to behave as the SM-like Higgs boson and to have a mass in the
range 
\begin{eqnarray}
122 \mbox{ GeV } \le m_h \le 128 \mbox{ GeV} \;, \label{eq:masswindow}
\end{eqnarray}
when including the two-loop corrections at ${\cal
  O}((\alpha_t+\alpha_\lambda+\alpha_\kappa)^2+\alpha_t \alpha_s)$ in
the default mixed $\DRbar$-OS scheme specified above and with OS
renormalisation in the top/stop and charged Higgs boson sectors as well
as an infrared mass regulator $M_R$ with $M_R^2 = 10^{-3}\mu_R^2$ to
treat the Goldstone problem. For details, we refer to \cite{Dao:2021khm}.  
The SM input values have been chosen
as~\cite{PhysRevD.98.030001,Dennerlhcnote}    
\begin{equation}
\begin{tabular}{lcllcl}
\quad $\alpha(M_Z)$ &=& 1/127.955\,, &\quad $\alpha^{\MSbar}_s(M_Z)$ &=&
0.1181\,, \\
\quad $M_Z$ &=& 91.1876~GeV\,, &\quad $M_W$ &=& 80.379~GeV\,, \\
\quad $m_t$ &=& 172.74~GeV\,, &\quad $m^{\MSbar}_b(m_b^{\MSbar})$ &=& 4.18~GeV\,, \\
\quad $m_c$ &=& 1.274~GeV\,, &\quad $m_s$ &=& 95.0~MeV\,,\\
\quad $m_u$ &=& 2.2~MeV\,, &\quad $m_d$ &=& 4.7~MeV\,, \\
\quad $m_\tau$ &=& 1.77682~GeV\,, &\quad $m_\mu$ &=& 105.6584~MeV\,,  \\
\quad $m_e$ &=& 510.9989~keV\,, &\quad $G_F$ &=& $1.16637 \cdot 10^{-5}$~GeV$^{-2}$\,.
\end{tabular}
\end{equation} 
In accordance with the SLHA format the soft SUSY
breaking masses and trilinear couplings are understood as $\DRb$ parameters at 
the scale
\begin{eqnarray} 
\mu_0 = M_{\text{SUSY}}= \sqrt{m_{\tilde{Q}_3} m_{\tilde{t}_R}} \;.  \label{eq:renscale}
\end{eqnarray}
This is also the renormalisation scale that we use in the computation
of the higher-order corrections. The scan ranges of our input
parameters are given in Tab.~\ref{tab:scanranges}. Note, that both
$\lambda$ and $\kappa$ are required to remain below 0.7 in order to roughly
ensure perturbativity below the GUT scale. Also $\lambda$, $\kappa$, $\mueff$ and
$\tan\beta$ are understood to be $\DRbar$ parameters at the scale
$M_{\text{SUSY}}$ according to the SLHA format. For the scan we kept
all CP-violating phases equal to zero.
\begin{table}[t]
\centering
\begin{minipage}[t]{0.5\textwidth}
\centering
    \begin{tabular}[t]{ll}
        parameter              & scan range [TeV]                \\ \hline
$M_{H^\pm}$            & [0.5, 1]               \\
$M_1,M_2$              & [0.4, 1]                    \\
$M_3$                  & 2               \\
$\mueff$               & [0.1, 1]              \\
$m_{\tilde{Q}_3}, m_{\tilde{t}_R}$      & [0.4, 3]               \\
$m_{\tilde{X}\neq \tilde{Q}_3,\tilde{t}_R}$      & 3        
\end{tabular}
\end{minipage}\hfill
\begin{minipage}[t]{0.5\textwidth}
\centering
    \begin{tabular}[t]{ll}
parameter              & scan range                 \\ \hline
$\tan\beta$            & [1, 10]                    \\
$\lambda$              & [0.01, 0.7]                 \\
$\kappa$               & $\lambda\cdot \xi$          \\
$\xi$                  & [0.1, 1.5]                   \\
$A_t$                  & [$-3$, 3] TeV                 \\
$A_{i\neq t}$          & [$-2$, 2] TeV 
\end{tabular}
\end{minipage}
\caption{Scan ranges for the random scan over the NMSSM parameter
  space, with $\tilde X = \tilde b_R, \tilde L, \tilde \tau$ and $i =
  b, \tau, \kappa$. Values of $\kappa=\lambda\cdot\xi>0.7$ are omitted.}
\label{tab:scanranges}
\end{table}
We neglected parameter points with any of the following mass configurations,
\begin{align}
 (i)\quad   &
 m_{\chi_i^{(\pm)}},m_{h_i}>\unit[1]{TeV},~ m_{\tilde{t}_2}>\unit[2]{TeV} \nonumber\\
 (ii)\quad  &  m_{h_i}-m_{h_j}<\unit[0.1]{GeV},~ m_{\chi_i^{(\pm)}}-m_{\chi_j^{(\pm)}}<\unit[0.1]{GeV} \nonumber\\
 (iii)\quad &  m_{\chi^\pm_1}<\unit[94]{GeV},~  m_{\tilde{t}_1}<\unit[1]{TeV}\nonumber \;.
\end{align}
With the first condition $(i)$ we avoid large logarithms in our fixed-order
calculation. The second condition $(ii)$
omits degenerate mass configurations for which the two-loop
part of the \NMSSMCALC code is not yet optimised. 
The third condition $(iii)$ takes into account model-independent lower 
limits for the lightest chargino and stop masses. 

\section{Investigation of Specific Benchmark Points \label{sec:bp2os}}
In the following, we present results for two benchmark points. One
point is the benchmark point {\tt P2OS} from our investigation of the
Higgs mass corrections at 
${\cal O} (\alpha_{\lambda\kappa}^2) \equiv {\cal
  O}((\alpha_t+\alpha_\lambda+\alpha_\kappa)^2+\alpha_t  
\alpha_s)$ in \cite{Dao:2021khm}. The other point is the benchmark
point {\tt BP10} of Ref.~\cite{Abouabid:2021yvw}. 
They have been chosen such that the SM-like Higgs boson
mass complies with 
our required mass window Eq.~(\ref{eq:masswindow}) at ${\cal O}
(\alpha_{\lambda\kappa}^2)$ when we choose OS renormalisation in the top/stop
sector. The charged Higgs mass here and in all other results presented in
the following is renormalised OS. The first parameter point {\tt P2OS}
features a large singlet admixture to the $h_u$-like mass and is
defined by the following input parameters: \s

\noindent
{\bf Parameter Point {\tt P2OS}:} All complex phases are set to zero
and the remaining input parameters are given by
\begin{eqnarray}
      |\lambda| &=& 0.59 \,, \; |\kappa|=0.23 \,, \; \mbox{Re}
              (A_\kappa) = -546 \mbox{ GeV}\,, \; |\mu_{\text{eff}}| =
              397 \mbox{ GeV} \,, \; \tan\beta = 2.05 \,, \nonumber \\
M_{H^\pm} &=& 922 \mbox{ GeV} \,,\; 
m_{\tilde{Q}_3} =1.2 \mbox{ TeV} \,, \; m_{\tilde{t}_R}= 1.37 \mbox{ TeV}
  \,, \; m_{\tilde{X}\neq \tilde{Q}_3,\tilde{t}_R}= 3 \mbox{ TeV}
  \,,\; \label{eq:P1OSnew} \\
A_t&=&-911 \mbox{ GeV} \,, \;  A_{i\neq t,\kappa}=0 \mbox{ GeV} \,, \;
|M_1| =656 \mbox{ GeV} \,, \; |M_2|= 679 \mbox{ GeV} \,,\; M_3= 2
\mbox{ TeV} \;. \nonumber
\end{eqnarray}
We apply the SLHA format in which $\mu_{\text{eff}}$ is taken as input
parameter. From this we compute $v_s$ by using
Eq.~(\ref{eq:mueffcalc}) ($\varphi_s$ is set to zero). \s

Since the trilinear Higgs self-couplings and the mass values are
closely related through the Higgs potential a discussion of the 
higher-order corrections to the trilinear Higgs self-couplings should
be completed by the information on the Higgs mass corrections. 
In  Table~\ref{tab:massvalues5} we hence give the mass values obtained for
{\tt P2OS} at tree level, at one-loop order and at two-loop level at
${\cal O}(\alpha_t \alpha_s)$, ${\cal O}(\alpha_t
(\alpha_s+\alpha_t))$ and the latest computed two-loop order ${\cal
  O}(\alpha^2_{\lambda\kappa})$  for OS renormalisation in the top/stop
sector, and in round brackets those for $\DRb$
renormalisation in the top/stop sector. Note that the numbers slightly
changed compared to those given in \cite{Dao:2021khm} due to a bug in
the VEV counterterm. The changes are in the sub percentage level. \s

In the table we also list in square brackets the main
singlet/doublet and scalar/pseudoscalar component of each mass
eigenstate. At ${\cal O}(\alpha_{\lambda\kappa}^2)$ the lightest Higgs
boson $h_1$ obtains a mass of around 125.3~GeV. Since it is $h_u$-like
it couples maximally to top quarks so 
that the LHC Higgs signal strengths are reproduced and it hence
behaves SM-like. In the following plots we will always label the Higgs
bosons according to their dominant admixture\footnote{They are mass
  eigenstates, however. The labeling only refers to the nature of
  these mass eigenstates.}, as this determines the
Higgs coupling strengths and consequently the size of the loop
corrections. This allows us to consistently compare and interpret the
impact of the loop corrections. \s
\begin{table}[t]
\begin{center}
 \begin{tabular}{|l||c|c|c|c|c|}
\hline
                                                    & ${h_1}$
                                                      \textcolor{red}{[$h_u$]}
   & ${h_2}$ \textcolor{blue}{[$h_s$]} & ${h_3}$ $[h_d]$& ${a_1}$ $[a_s]$ & ${a_2}$ $[a_d]$\\ \hline \hline
tree-level  & \textcolor{red}{96.86}     & \textcolor{blue}{112.10}  &  926.25  & 511.34  & 925.86 \\ \hline
one-loop & \textcolor{blue}{129.01}    & \textcolor{red}{135.09}  &
                                                                    926.69   & 512.55   & 925.08  \\
& \textcolor{red}{(116.3)}  & \textcolor{blue}{(130.1)}  & (926.33)   & (512.66)   & (925.18)   \\ \hline
two-loop  ${\cal O}(\alpha_t \alpha_s)$ & \textcolor{red}{121.36}    &
\textcolor{blue}{129.7}  & 926.37   & 512.62   & 925.11  \\
 & \textcolor{red}{(121.65)}  & \textcolor{blue}{(130.39)}  & (926.46)   & (512.61)   & (925.15)  \\ \hline
two-loop ${\cal O}(\alpha_t(\alpha_s+ \alpha_t))$  & \textcolor{red}{126.09}    & \textcolor{blue}{130.04}  & 926.49   & 512.62   & 925.11  \\
& \textcolor{red}{(121.54)}    & \textcolor{blue}{(130.38)}  & (926.45)   & (512.61)   & (925.15)  \\ \hline
two-loop ${\cal O}(\alpha_{\lambda\kappa}^2)$ & \textcolor{red}{125.25}  & \textcolor{blue}{129.91}  & 926.62   & 511.91   & 925.07  \\
& \textcolor{red}{(121.67)}    & \textcolor{blue}{(130.20)}  & (926.52)   & (512.12)   & (925.14)   \\ \hline
 \end{tabular}
 \caption{{\tt P2OS}: Mass values in GeV and main components of the neutral Higgs
  bosons at tree-level, one-loop, two-loop $\order{\alpha_t\alpha_s}$,
  two-loop $\order{\alpha_t(\alpha_s + \alpha_t)}$ and at two-loop
  $\order{\alpha_{\lambda\kappa}^2}$
  obtained by using OS ($\overline{\mbox{DR}}$)  renormalisation in the
  top/stop sector. \textcolor{red}{Red} numbers relate to states that are dominantly
  \textcolor{red}{$h_u$}-like, \textcolor{blue}{blue} numbers relate
  to dominantly \textcolor{blue}{$h_s$}-like states.}
\label{tab:massvalues5}
\end{center}
\end{table}

The second parameter point {\tt BP10} features a resonantly enhanced
Higgs pair production cross section in gluon fusion and is given by:\s

\noindent
{\bf Parameter Point {\tt BP10}:} All complex phases are set to zero and
the remaining input parameters are given by 
\begin{eqnarray}
      |\lambda| &=& 0.65 \,, \; |\kappa|=0.65 \,, \; \mbox{Re}
              (A_\kappa) = -432 \mbox{ GeV}\,, \; |\mu_{\text{eff}}| =
              225 \mbox{ GeV} \,, \; \tan\beta = 2.6 \,, \nonumber \\
M_{H^\pm} &=& 611 \mbox{ GeV} \,,\; 
m_{\tilde{Q}_3} =1304 \mbox{ GeV} \,, \; m_{\tilde{t}_R}= 1576 \mbox{ GeV}
  \,, \; m_{\tilde{X}\neq \tilde{Q}_3,\tilde{t}_R}= 3 \mbox{ TeV}
  \,,\; \nonumber \\
A_t&=&46 \mbox{ GeV} \,, \;  A_{b}=-1790 \mbox{ GeV} \,, \;
A_\tau = -93 \mbox{ GeV} \,, \; A_c = 267 \mbox{ GeV} \,, \nonumber \\
A_s &=& -618 \mbox{ GeV} \,, \; A_\mu = 1851 \mbox{ GeV} \,,\;
A_u = -59 \mbox{ GeV} \,, \; A_d = -175 \mbox{ GeV} \,,\nonumber \\
A_e &=& 1600 \mbox{ GeV} \,,\;
|M_1| = 810 \mbox{ GeV} \,,\; |M_2|= 642 \mbox{ GeV} \,,\; M_3= 2
        \mbox{ TeV} \;. \label{eq:P3OS} 
\end{eqnarray}
The mass values that are
obtained at the different loop levels are summarised in
Tab.~\ref{tab:massvaluesbp10} for OS ($\overline{\mbox{DR}}$)
renormalisation in the top/stop sector. \s
\begin{table}[t]
\begin{center}
 \begin{tabular}{|l||c|c|c|c|c|}
\hline
 & ${h_1}$ $[h_u]$  & ${h_2}$ $[h_s]$ & ${h_3}$ $[h_d]$ & ${a_1}$ $[a_s]$ & ${a_2}$ $[a_d]$ \\ \hline \hline
tree-level  & 97.21  &  307.80 & 626.13 & 556.71 & 617.22   \\ \hline
one-loop & 131.46    &  299.65  & 625.96 & 543.58  &  615.82 \\
& (114.81)    & (299.28)  & (625.52)  & (543.69)  & (616.01) \\ \hline
two-loop  ${\cal O}(\alpha_t \alpha_s)$ & 118.90  & 299.40 & 625.78 & 543.73 &
615.90 \\ 
& (120.36) & (299.38) & (625.58) &  (543.60) & (615.96) \\
\hline
two-loop ${\cal O}(\alpha_t(\alpha_s+ \alpha_t))$ & 123.53 & 299.44 &  625.89 & 543.73 & 615.90 \\
& (120.14)  & (299.38) & (625.57) & (543.60) & (615.96)    
\\ \hline
two-loop ${\cal O}(\alpha_{\lambda\kappa}^2)$ & 122.36 & 300.27 &  625.94  & 543.34 &  615.91 \\
& (119.97) & (299.90) & (625.65) & (543.47) & (616.01)\\ \hline
 \end{tabular}
 \caption{{\tt BP10}: Mass values in GeV and main components of the neutral Higgs
  bosons at tree-level, one-loop, two-loop $\order{\alpha_t\alpha_s}$,
  two-loop $\order{\alpha_t(\alpha_s + \alpha_t)}$ and at two-loop
  $\order{\alpha_{\lambda\kappa}^2}$
  obtained by using OS ($\overline{\mbox{DR}}$) renormalisation in the top/stop sector.}
\label{tab:massvaluesbp10}
\end{center}
\end{table}

The impact of the loop corrections on the Higgs boson masses has been
discussed extensively in \cite{Dao:2021khm}. Let us therefore here
state only the main features. The $h_u$-like tree-level Higgs mass value changes
considerably when one-loop corrections are included, with a smaller
change in the $\DRbar$ scheme, as in this scheme we 
already partly resum higher-order corrections. The relative ${\cal O}(\alpha_t
\alpha_s)$ corrections compared to the one-loop result are at the
several per-cent level and move the obtained mass values in the two
renormalisation schemes closer to each other, whereas the additional
inclusion of the ${\cal O}(\alpha_t^2)$ corrections increases the
difference again (in the OS scheme). The newest corrections at ${\cal
  O}(\alpha_{\lambda\kappa}^2)$ move the two values a little bit closer again.


\subsection{Impact on the Effective Trilinear Higgs Self-Coupling}

\begin{figure}[t!]
    \centering
\includegraphics[width=0.49\textwidth]{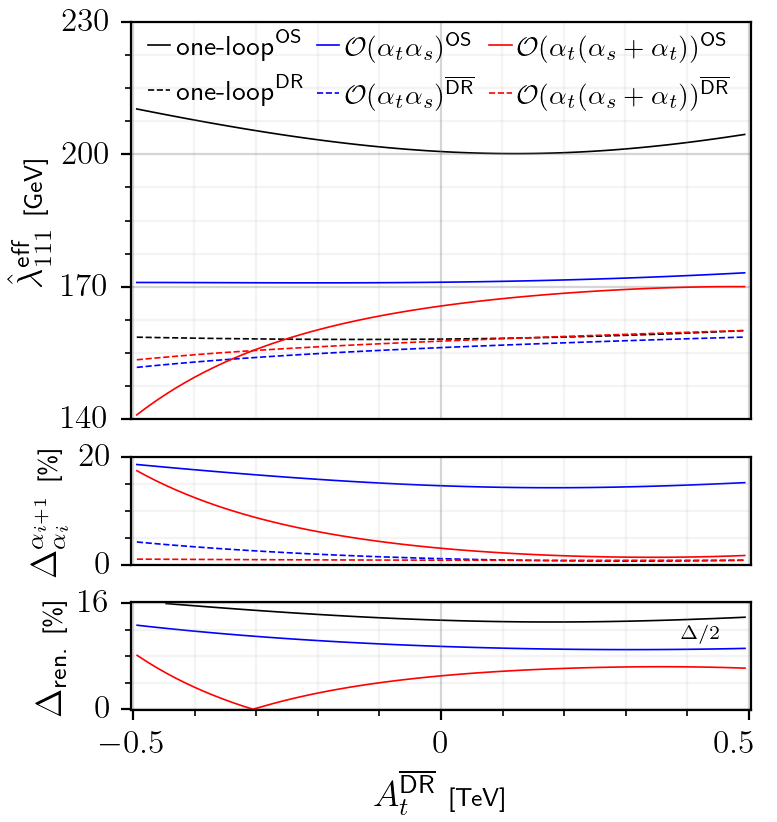}
\includegraphics[width=0.49\textwidth]{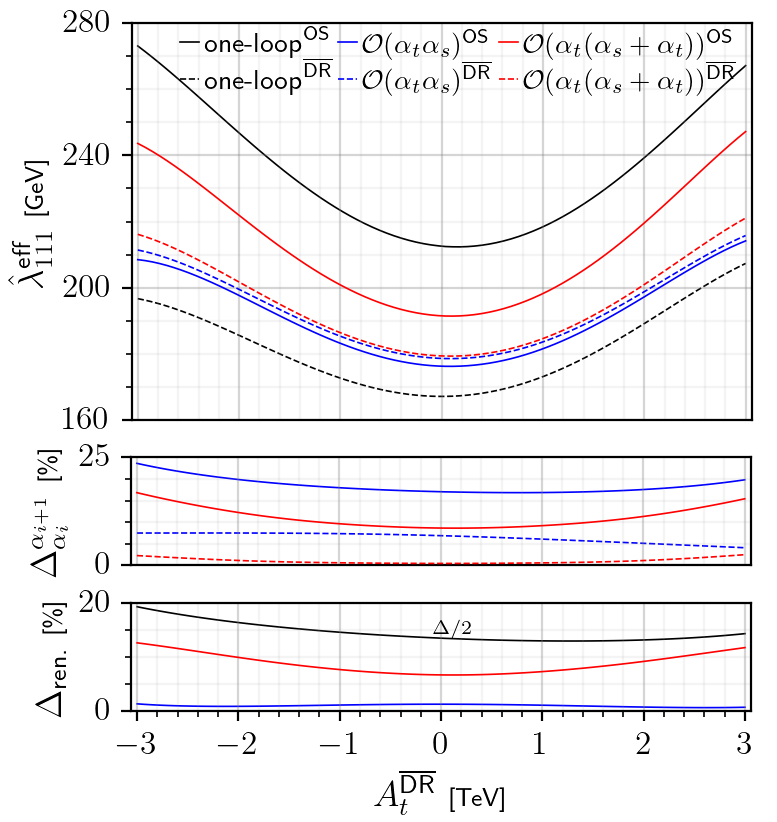}
    \caption{Upper: Effective trilinear coupling
      $\hat\lambda^{\text{eff}}_{111}$ of the $h_u$-dominated Higgs
      mass eigenstate as a function 
      of $A_t^{\overline{\text{DR}}}$ for {\tt P2OS} (left) and {\tt BP10} (right) at
      one-loop order (black), two-loop ${\cal O}(\alpha_t \alpha_s)$
      (blue) and two-loop ${\cal O}(\alpha_t (\alpha_s + \alpha_t))$ (red)
      in the OS (full) and $\DRb$ scheme (dashed). Middle: The
      relative correction as defined in Eq.~(\ref{eq:corrdef}). Lower: The
      relative renormalisation scheme dependence as defined in
      Eq.~(\ref{eq:rendelta}) for all three loop orders. The label
      $\Delta/2$ refers to the black line. \label{fig:trilinearp12os}}
\end{figure}

In Fig.~\ref{fig:trilinearp12os} (left) we present for the parameter point {\tt
  P2OS} with the large singlet admixture the effective trilinear Higgs
self-coupling $\hat\lambda^{\text{eff}}_{111}$, as defined in
Eq.~(\ref{eq:effdef}), of the dominantly 
$h_u$-like Higgs boson for OS (full) and $\overline{\mbox{DR}}$
(dashed) renormalisation in the top/stop sector as a function of the stop
trilinear coupling $A_t$. The dominantly $h_u$-like
  Higgs boson here always is the lightest mass eigenstate.
Note, that here and in the following the
$A_t$ is always the $\overline{\mbox{DR}}$
value.\footnote{The corresponding value of
    $A_t^{\text{OS}}$ differs by 0-20\% from $A_t^{\overline{\text{DR}}}$ such that the
    overall shape of the plots remains the same.} 
Shown are the results at one-loop order
(black), two-loop ${\cal O}(\alpha_t \alpha_s)$ (blue) and at the
newly calculated two-loop ${\cal O}(\alpha_t (\alpha_s +
\alpha_t))$ (red). Note that the loop-corrected rotation matrix ${\cal
R}^{l,\text{eff}}$ for the rotation to the mass eigenstates is taken
consistently at the respective loop order. 
We plot here only the variation of $A_t$ between $-500$ and $+500$~GeV. In
this region the phenomenology at ${\cal O} (\alpha_t (\alpha_s +
\alpha_t))$ is in accordance with the LHC Higgs
data.\footnote{Different higher-order corrections to the SM-like Higgs
  boson mass obviously imply different mass values and mixing angles
  and hence affect the compatibility with the Higgs data.} \s

The steep decrease of the full red curve towards 
negative $A_t$ values is due to the approach to the cross-over point
where the singlet-like and doublet $h_u$-like Higgs state interchange their
roles with respect to their mass ordering. This point is 
located outside of the shown region in the plot, at $A_t=-900$~GeV, {\it cf.}~Fig.~3
in Ref.~\cite{Dao:2021khm}. The interchange of the roles of the two
lightest mass eigenstates takes place due to the large singlet
admixture for this parameter point which induces the transition
between the $h_s$- and $h_u$-like 
interaction state. 
For the trilinear coupling $\hat{\hat
  \lambda}_{ijk}$ in the interaction basis after singling out the Goldstone boson, {\it
  cf.}~Eq.~(\ref{eq:singling}), this of course does not occur.
The multiplication with the mixing 
matrices ${\cal R}^{l,\text{eff}}$ causes the mixing of the singlet
and doublet states and also mixes in higher orders, as we do not
evaluate the mixing matrix multiplication strictly at the considered
loop order. \s
%

In order to quantify the 
impact of the new additional corrections we define -- for a given renormalisation
scheme -- the relative change in the trilinear coupling value
when going from loop order $\alpha_i$ to the loop order $\alpha_{i+1}$,
which includes the next level of corrections, as
\beq
\Delta^{\alpha_{i+1}}_{\alpha_i} = \frac{\left| \lambda^{\alpha_{i+1}}
    - \lambda^{\alpha_i} \right|}{\lambda^{\alpha_i}} \;.  
\label{eq:corrdef}
\eeq  
The relative corrections amount to $\Delta^{\alpha_t
  \alpha_s}_{\text{one-loop}}=14$--$19$\% in the OS scheme when we
include the ${\cal O}(\alpha_t \alpha_s)$ 
corrections beyond one-loop order. When we include the ${\cal
  O}(\alpha_t(\alpha_s+\alpha_t))$ corrections in addition to the
available two-loop ${\cal O}(\alpha_t \alpha_s)$ corrections the
relative change is smaller with $\Delta^{\alpha_t
  (\alpha_s+\alpha_t)}_{\alpha_t \alpha_s}=1.5$--$18$\%. 
As expected the relative change decreases
with increasing higher-order in the corrections. Note, that the
relative corrections can be positive or negative. 
In the $\overline{\mbox{DR}}$ scheme the corrections are much
smaller. We have for the 
relative ${\cal O}(\alpha_t \alpha_s)$ corrections compared to
one-loop order 0.8--4\%, and the new corrections
change the coupling by about 1\%. 
The reason is that the $\overline{\mbox{DR}}$ scheme already partly resums
higher-order corrections. 
\s

The lower panel in
Fig.~\ref{fig:trilinearp12os} shows the 
relative change in the corrections to the trilinear Higgs
self-coupling at fixed loop order when we change the renormalisation
scheme in the top/stop sector,\footnote{Note that we have consistently
converted the $A_t^{\overline{\text{DR}}}$ input parameter to the
$A_t^{\text{OS}}$ value when we change the top/stop renormalization
scheme from $\overline{\text{DR}}$ to OS.} 
\begin{equation}
\Delta_{\text{ren}} = \frac{\left| \lambda^{m_t(\DRbar)}-
    \lambda^{m_t(\text{OS})}\right| }{ \lambda^{m_t(\DRbar)} } \;. \label{eq:rendelta}
\end{equation} 
The comparison of the results in the two different
renormalisation schemes can be used to estimate the uncertainty
on the trilinear Higgs self-coupling due to missing higher-order
corrections. As expected, the renormalisation scheme
dependence is reduced when more higher-order
corrections are included. The renormalisation
scheme dependence continuously decreases from 
one-loop order with 26--33\%, to 9--13\% at
${\cal O}(\alpha_t \alpha_s)$ and to 0.01--8\% at ${\cal
  O}(\alpha_t (\alpha_s+\alpha_t))$. \s

In Fig.~\ref{fig:trilinearp12os} (right) we show our results for the benchmark point {\tt
  BP10}. For this point, the relative corrections in the OS scheme are
slightly larger than for {\tt P2OS}. We have $\Delta^{\alpha_t
  \alpha_s}_{\text{one-loop}}=17$--$24$\% and $\Delta^{\alpha_t
  (\alpha_s+\alpha_t)}_{\alpha_t \alpha_s}=9$--$17$\%. In the
$\overline{\mbox{DR}}$ scheme the corrections are smaller, the
relative ${\cal O}(\alpha_t \alpha_s)$ corrections compared to
one-loop order are of 4--7\%, and the new corrections
change the coupling by 0.4--2\%. The renormalisation
scheme dependence decreases from 
one-loop order with 26--39\% to 0.7--1.4\% at
${\cal O}(\alpha_t \alpha_s)$. The scheme dependence slightly increases
again at ${\cal O}(\alpha_t (\alpha_s+\alpha_t))$ where it is
7--13\%. This is
a behaviour that we already observed in the loop corrections to the
Higgs boson masses \cite{Dao:2019qaz}.\footnote{Incomplete two-loop
corrections cannot necessarily be expected to reduce the uncertainty
when including further corrections as there might be missing
cancellations. The complete two-loop corrected results, however,
should reduce the renormalisation scheme dependence compared to the
complete one-loop result in a perturbative expansion in the coupling constants.}\s

In summary, for both benchmark points the inclusion of the new
two-loop corrections has an impact of a few per cent and we find a
renormalisation scheme dependence of typical two-loop order. The
behaviour is similar to the one we found for the Higgs mass
corrections.

\begin{figure}[h!]
    \centering
    \includegraphics[width=0.55\textwidth]{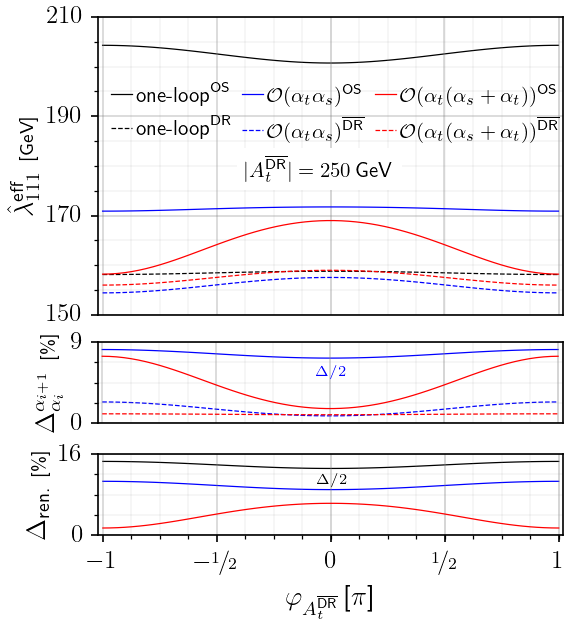}\\
    \caption{Effective Higgs self-coupling
      $\hat\lambda^{\text{eff}}_{111}$ of the $h_u$-dominated Higgs
      mass eigenstate for {\tt P2OS} as a 
      function of $\varphi_{A_t}$ for $|A_t^{\overline{\text{DR}}}|=250$~GeV. Color and line
      codes are the same as in Fig.~\ref{fig:trilinearp12os}. \label{fig:cpviol}}
\end{figure}

\paragraph{CP violation}
In Fig.~\ref{fig:cpviol} we show for the parameter point {\tt P2OS}
the loop corrections to the effective trilinear Higgs
self-coupling $\hat{\lambda}^{\text{eff}}_{111}$ of the dominantly
$h_u$-like Higgs boson as a function of the CP-violating phase
$\varphi_{A_t}$ of $A_t^{\overline{\text{DR}}}$. The colour and
line codes are the same as in Fig.~\ref{fig:trilinearp12os}. In order to
avoid too large singlet-doublet mixing effects we chose
$|A_t^{\overline{\text{DR}}}|=250$~GeV. 
CP violation due to the phase $\varphi_{A_t}$ is a loop-induced effect.
Since $A_t$ enters at one-loop level,
  the trilinear Higgs self-coupling shows a dependence on the
  CP-violating phase, which for this parameter point turns out to be
  larger in the OS than in the $\overline{\mbox{DR}}$ renormalisation
  scheme. The almost flat dependence on the CP-violating phase of the
  OS curve at ${\cal O}(\alpha_t \alpha_s)$ is due to accidental
  cancellations which we explicitly checked. At ${\cal
    O}(\alpha_t(\alpha_s+\alpha_t))$, we see a stronger dependence
  again. In the $\overline{\mbox{DR}}$ scheme both two-loop orders
  show about the same dependence on the phase $\varphi_{A_t}$.

\subsection{Impact on the Higgs-to-Higgs Decays}
\begin{figure}[t!]
    \centering
    \includegraphics[width=0.495\textwidth]{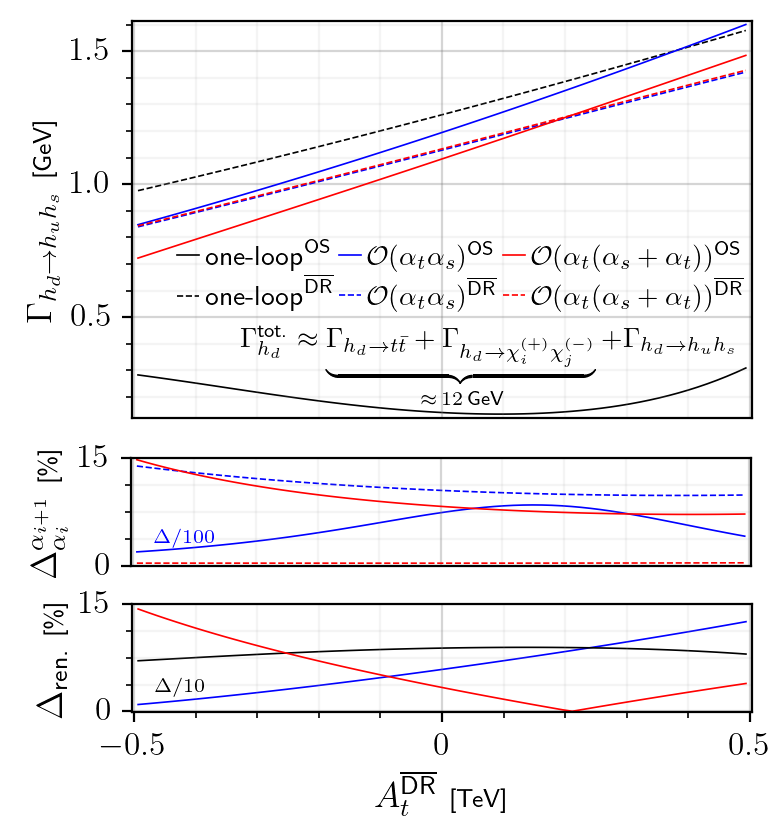}
    \includegraphics[width=0.495\textwidth]{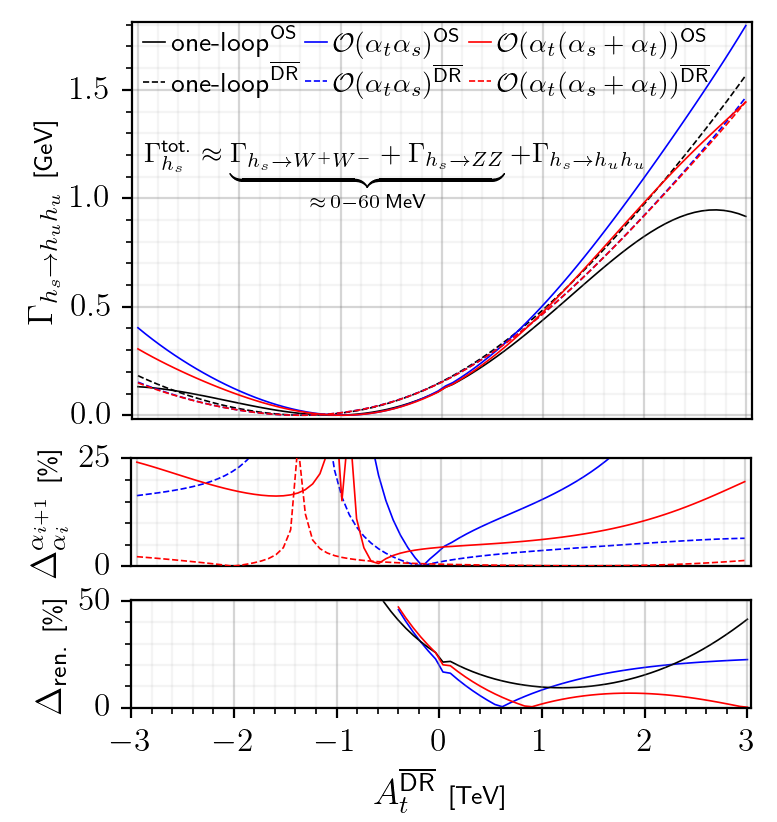}
    \caption{Upper: Partial decay width of $h_d \to h_u h_s$ for {\tt P2OS}
      (left) and of $h_s \to h_u h_u$ for {\tt BP10} (right) at
      one-loop order (black), two-loop ${\cal O}(\alpha_t \alpha_s)$
      (blue) and two-loop ${\cal O}(\alpha_t (\alpha_s + \alpha_t))$ (red)
      in the OS (full) and $\DRb$ scheme (dashed) as a function of
      $A_t^{\overline{\text{DR}}}$. In the right plot
      the dashed blue and red lines nearly lie on top of each other. Middle: The
      relative correction defined analogously to
      Eq.~(\ref{eq:corrdef}), but for the partial decay width. Lower: The
      relative renormalisation scheme dependence defined analogously
      to Eq.~(\ref{eq:rendelta}), but for the partial decay width, for
      all three loop orders.\label{fig:decay} }
\end{figure}

We now turn to the impact of the computed higher-order corrections on
the partial decay widths for Higgs-to-Higgs decays. 
The decay width for the Higgs decay $h_i$ into a Higgs pair $h_j h_k$
is given by
\beq
\Gamma (h_i \to h_j h_k) =
\frac{\beta^{1/2}(M_{h_i}^2,M_{h_j}^2,M_{h_k}^2)}{16 \pi
  (1+\delta_{jk}) M_{h_i}^3} |{\cal M}_{h_i \to h_j h_k}|^2 \;,
\eeq
where $\beta(x,y,z) = (x-y-z)^2 - 4yz$ is the two-body phase space function
and the decay amplitude ${\cal M}_{h_i \to h_j h_k}$ is calculated according to
Eq.~(\ref{eq:decayamplitude}). 
We show in Fig.~\ref{fig:decay} (left) the partial decay width of the
doublet-like CP-even Higgs boson $h_d$ into a pair of a SM-like Higgs boson 
$h_u$ and a singlet-dominated Higgs $h_s$, $\Gamma (h_d \to h_u h_s)$, at
one-loop level and at two-loop ${\cal O}(\alpha_t \alpha_s)$ and
${\cal O}(\alpha_t (\alpha_s + \alpha_t))$ for {\tt P2OS}, as a
function of $A_t$.\footnote{Note that, as stated above, the notation for the Higgs
  states only relates to their dominant component, but still they are mass
and not interaction eigenstates.} This decay
is the largest of the Higgs-to-Higgs decays for this parameter
point. For {\tt BP10} the largest one is given by $h_s \to h_u h_u$
which we show in Fig.~\ref{fig:decay} (right). This is also the
resonant contribution that increases the production process of an $h_u
h_u$ Higgs pair which we will discuss later. 
We include both the {\bf Z} matrix of \eqref{eq:zwfrmatrix} and the
Higgs mass values calculated at the corresponding same 
loop order as the one for which we calculate the higher-order corrections to the
trilinear Higgs self-coupling. This of course also implies that the
kinematical factor in the decay amplitude changes with the loop
order. For both parameter points, we observe a reduction of both the 
relative correction and the renormalisation scheme dependence when we
move from one- to two-loop order with the effect being less pronounced
in the $\DRb$ than in the OS scheme as the former already partly
resums higher-order corrections. \s

For {\tt P2OS} the relative corrections for the partial decay width in
the OS scheme amount to more than 100\% when including the ${\cal
  O}(\alpha_t \alpha_s)$ corrections in addition to the one-loop
corrections. The reason is the small one-loop decay width. In the
$\overline{\mbox{DR}}$ scheme the relative corrections amount to 
10--14\%. The relative effect of the new two-loop
corrections is much less as expected and reaches $\Delta^{\alpha_t 
  (\alpha_s+\alpha_t)}_{\alpha_t
  \alpha_s}=7$--$15$\% (0.4\%) in the OS
($\overline{\mbox{DR}}$) scheme. The renormalisation scheme dependence
decreases from ${\cal O}(71$--$90\%)$ at one-loop level
to a maximum of 13\% at two-loop ${\cal O} (\alpha_t
\alpha_s)$ and at most 14\% at two-loop ${\cal O}(\alpha_t 
  (\alpha_s+\alpha_t))$.
For the benchmark point {\tt BP10} we find that for most $A_t$ values
the relative corrections of our new two-loop corrections are smaller
compared to the relative corrections when moving from one- to two-loop
order. Note that we cut some of the lines in the middle plot of
Fig.~\ref{fig:decay} (right) as here the relative corrections become
artificially large due to comparatively very small widths at the previous loop order.
The reduction in the renormalisation scheme dependence when
moving from one- to two-loop order is less obvious as can be seen from the lower
panel in Fig.~\ref{fig:decay} (right). 
The renormalisation scheme dependence
becomes artificially large here where the $\overline{\mbox{DR}}$ result for
the partial decay width is very small. \s 

In both scenarios the partial decay widths can be become as large as
about 1.7~GeV. In {\tt P2OS} this leads to a branching ratio of about
12\% at most, taking into account the dominant decay channels. In {\tt
  BP10} we get a maximum branching ratio of more than 70\%. \s

Our results show that the higher-order corrections to the
decay width have a substantial impact in particular when moving from
one-loop to two-loop order. Furthermore, also at the two-loop level the
inclusion of the ${\cal O}(\alpha_t^2)$ corrections on top of the
available ${\cal O}(\alpha_t \alpha_s)$ corrections leads to
significant changes in the decay width both for {\tt P2OS} and {\tt
  BP10}. At the phenomenological level, the impact
of the changes depends on the relative size of the Higgs-to-Higgs
decay widths compared to the other decay widths.\footnote{The partial widths for the computation of the branching
  ratios are obtained from the code {\tt NMSSMCALC} \cite{Baglio:2013iia}. It includes the
  dominant higher-order QCD corrections and in the Higgs-to-Higgs
  decays the higher-order corrections up to ${\cal O}(\alpha_t
  (\alpha_s+\alpha_t))$.}

\section{Scatter Plots}
\label{sec:scatter}
After the investigation of two specific benchmark points we aim to
get an overall picture of the corrections by investigating scatter
plots. These plots contain all parameter scenarios that we obtained
from our scan and that comply with the included constraints described
above. 

\subsection{The Trilinear Higgs Self-Coupling}
\begin{figure}[t!]
    \centering
    \includegraphics[width=0.48\textwidth]{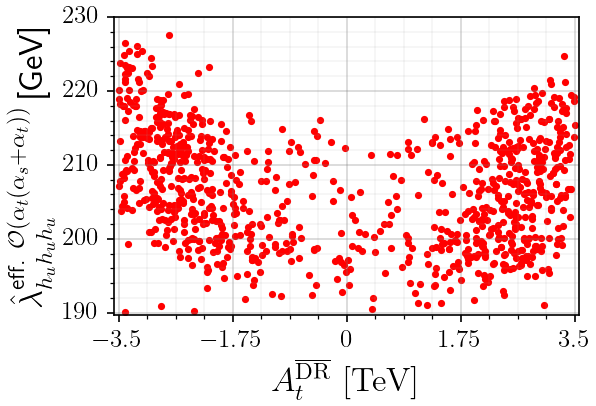}
\hspace*{-0.2cm}
    \includegraphics[width=0.48\textwidth]{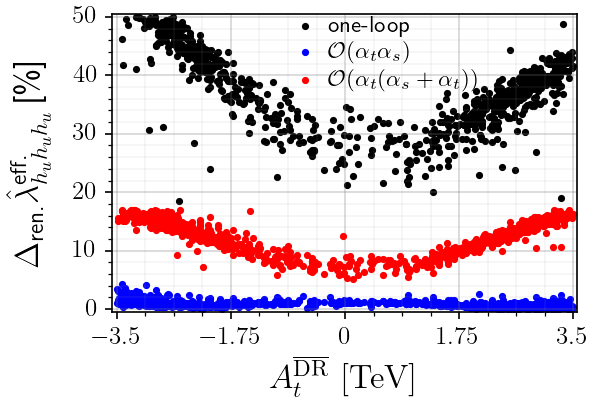}
    \caption{Left: The effective trilinear self-coupling
      $\hat\lambda^{\text{eff}}_{h_u h_u h_u}$ of the
      SM-like $h_u$-dominated Higgs mass eigenstate as
      a function of $A_t^{\overline{\text{DR}}}$ at ${\cal O}(\alpha_t (\alpha_s+\alpha_t))$ in
      the OS scheme. Right: The renormalisation scheme dependence as a
      function of $A_t^{\overline{{\text DR}}}$ at one-loop order (black), at two-loop ${\cal
        O}(\alpha_t \alpha_s)$ (blue) and at two-loop ${\cal O}(\alpha_t
      (\alpha_s+\alpha_t))$ (red).
\label{fig:trilinearscatter}}
\end{figure}

Figure~\ref{fig:trilinearscatter} (left) displays for all generated valid
parameter scenarios the effective trilinear Higgs self-coupling
$\hat\lambda^{\text{eff}}_{h_u h_u h_u}$ of the Higgs mass eigenstate that
is dominantly $h_u$-like, at ${\cal O} (\alpha_t (\alpha_s + \alpha_t))$
in the OS renormalisation scheme of the top/stop sector as a function
of $A_t \equiv A_t^{\overline{\text{DR}}}$. Note that,
depending on the parameter point, this is not necessarily always the
same Higgs mass eigenstate.
The right plot shows the renormalisation scheme dependence 
at the one- and considered two-loop orders. We see the same trend as
observed for the benchmark points. The scheme dependence at one-loop
order is rather large, varying between about 20\% and more than
50\%. It is considerably reduced upon inclusion of the two-loop ${\cal
O} (\alpha_t \alpha_s)$ corrections where it ranges between about 1
and 5\%. After the additional inclusion of the ${\cal O} (\alpha_t^2)$
corrections 
the scheme dependence increases again to values between 5 and 18\% and
reflects the necessity to include all corrections at a given loop
order in order to make a reliable statement on the scheme dependence. The
scheme dependence at two-loop order is well below the one at one-loop
level as expected. \s

Turning to the values of the trilinear Higgs
self-coupling $\hat\lambda^{\text{eff}}_{h_u h_u h_u}$ we find that it lies
between 190 and 228~GeV. For the 
SM-coupling we have 
\beq
\lambda_{HHH}^{\text{SM}} = \frac{3 M_H^2}{v} = 191~\mbox{GeV} \;,
\label{eq:smtrilvalue}
\eeq
for $M_H=125.09$~GeV and $v= \sqrt{\sqrt{2} G_F} \approx 246.22$~GeV. 
Taking into account the residual theoretical uncertainty at ${\cal O}
(\alpha_t (\alpha_s + \alpha_t))$ due to missing higher-order
corrections, the NMSSM $h_u$-like trilinear Higgs self-coupling
complies with the one found in the SM. This means that taking into account
the LHC Higgs data results which push the discovered Higgs boson very
close to the SM expectation, we find that the NMSSM $h_u$-like
trilinear Higgs self-coupling is also very SM-like once all dominant
higher-order corrections are taken into account. This has important
implications for the cross section values of Higgs pair production
(see our discussion in Sec.~\ref{sec:higgspair}). Note also that these values for
$\hat\lambda^{\text{eff}}_{h_u h_u h_u}$ lie well within the present
experimental limits on the SM trilinear Higgs self-coupling which are between $-0.4$
and $6.3$ times the SM value as reported by ATLAS
\cite{atlaspaperdihiggs} and between $-1.7$ and $8.7$ times the SM value
as found by CMS \cite{CMS:2022hgz}  (both assuming a SM-like
top-Yukawa coupling). 

\subsection{Correlation between Trilinear Higgs Self-Coupling and Mass}
\begin{figure}[t!]
    \centering
    \includegraphics[width=0.48\textwidth]{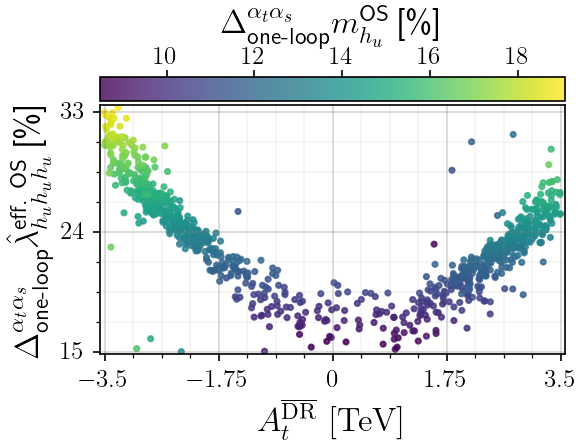}
    \includegraphics[width=0.48\textwidth]{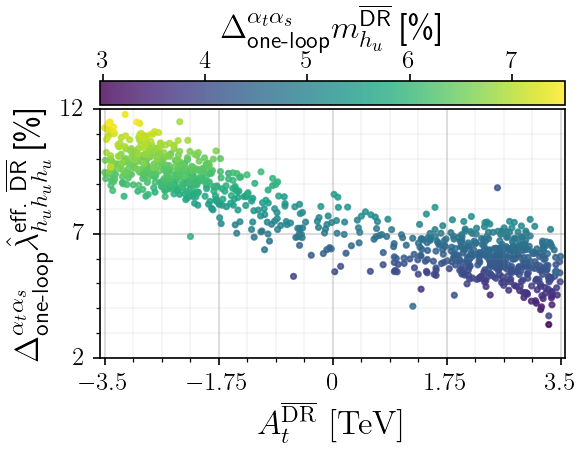} \\
    \includegraphics[width=0.48\textwidth]{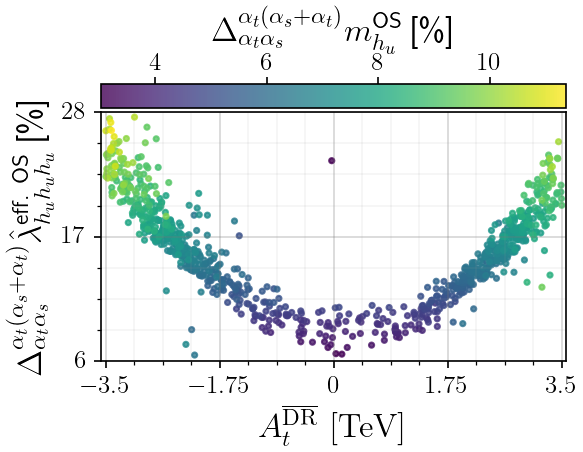}
     \includegraphics[width=0.48\textwidth]{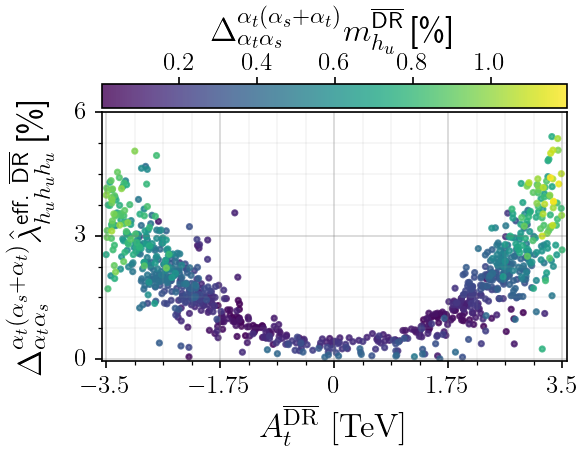}
 \caption{Relative size of the loop-corrected effective trilinear
   Higgs self-coupling of the $h_u$-like Higgs boson
   $\hat\lambda^{\text{eff}}_{h_u h_u h_u}$ w.r.t.~the next 
   lower order in the OS (left) and the $\overline{\mbox{DR}}$
   scheme (right) at 
   ${\cal O}(\alpha_t\alpha_s)$ (upper) and ${\cal O}
   (\alpha_t(\alpha_s+\alpha_t))$ (lower) as a function of
   $A_t^{\overline{\text{DR}}}$. The colour bar shows the 
   corresponding values for the $h_u$-like Higgs
   mass.
\label{fig:relsizescatter}}
\end{figure}

Figure~\ref{fig:relsizescatter} puts the relative corrections to the
effective trilinear Higgs self-coupling of the SM-like $h_u$-dominated
Higgs boson in relation to the relative corrections of its  
mass value. The
scatter plots of the valid parameter scenarios displayed in
Fig.~\ref{fig:relsizescatter} (upper) show that the relative
impact of the inclusion of the ${\cal O}(\alpha_t \alpha_s)$
corrections on top of the one-loop corrections is of about 15--35\% in
the OS and of roughly 3--12\% in the $\overline{\mbox{DR}}$ scheme for
the trilinear Higgs self-coupling. As for the masses, we find here
8--19\% in the OS and 3--8\% in the $\overline{\mbox{DR}}$ scheme. Both
corrections are correlated, larger corrections in the trilinear Higgs
self-coupling correspond to larger corrections for the Higgs mass. As
can be inferred from Fig.~\ref{fig:relsizescatter} (lower), the
relative size of the additional ${\cal O}(\alpha_t^2)$ corrections
amounts to about 5--27\% in the OS scheme and roughly 1--6\% in the
$\overline{\mbox{DR}}$ scheme for the trilinear Higgs self-coupling
which compares to 3--11\% in the OS scheme and 0.1--1.1\% in the
$\overline{\mbox{DR}}$ scheme for the masses. The mass corrections are
in general smaller than the corrections to the trilinear Higgs
self-couplings, and the corrections in the OS scheme are generally
larger than in the $\overline{\mbox{DR}}$ scheme which partly resums
higher-order corrections. \s

Overall we find for our parameter points compatible with the applied
constraints that the effective trilinear Higgs self-coupling values at ${\cal O}(\alpha_t
(\alpha_s + \alpha_t))$ are in general smaller in the
$\overline{\mbox{DR}}$ scheme compared to the OS scheme, as is the
case for the mass values. For both
schemes we see that the coupling values increase with increasing mass
values. This behavior reflects what we expect from the SM relation
Eq.~(\ref{eq:smtrilvalue}), and as stated above, within the residual
theoretical uncertainty the trilinear coupling values also comply with
the SM result.

\section{Higgs Pair Production \label{sec:higgspair}}
\begin{figure}[t]
  \centering
\includegraphics[width=0.4\textwidth,angle=90]{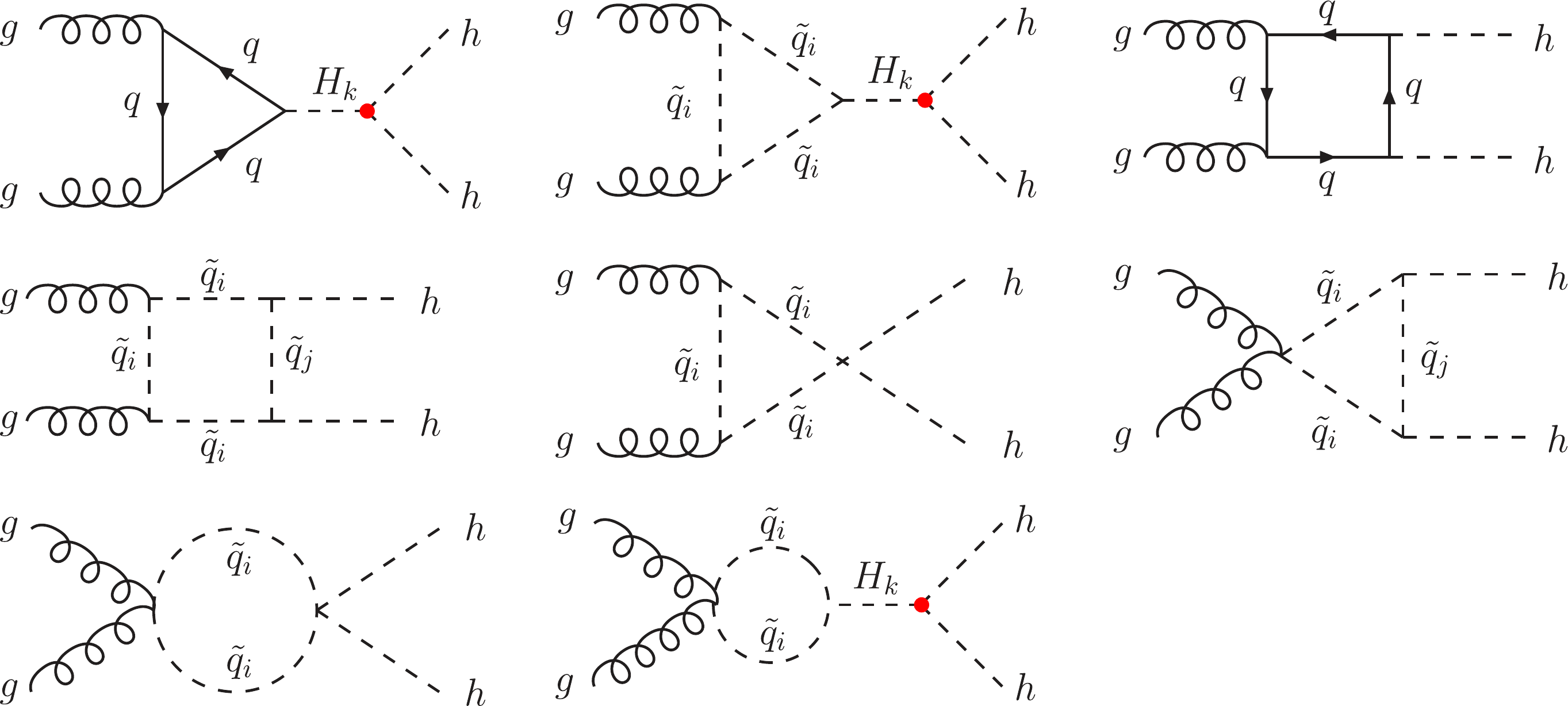}
  \caption{Generic diagrams contributing to pair production of a SM-like
    NMSSM Higgs boson $h$ in gluon fusion. The loops involve top and
    bottom (s)quarks, $q=t,b$, $\tilde{q}=\tilde{t},\tilde{b}$, $i,j=1,2$. The
    $s$-channel diagrams proceed via $H_k=H_1,H_2,H_3$, with one of
    these being the SM-like $h$, depending on the parameter
    choice.}
 \label{fig:doubleh}
\end{figure}

In this section we want to analyse what we can learn from our
higher-order results to the trilinear Higgs self-coupling about the
impact of the electroweak corrections on Higgs pair production. Higgs
pair production gives access to the trilinear Higgs self-coupling and
the measurement of the Higgs self-interactions provides the ultimate
test \cite{Djouadi:1999gv,Djouadi:1999rca,Muhlleitner:2000jj,DiMicco:2019ngk} of the
Higgs mechanism for the generation of particle masses. At the
LHC the dominant Higgs pair production process is given by gluon
fusion into Higgs pairs \cite{DiMicco:2019ngk,deFlorian:2016spz,Baglio:2012np}. The
loop-induced process is mediated by top-quark loops and by bottom-quark
loops, the latter contributing at the percent level. Higher-order QCD
corrections are important, increasing the cross section by roughly a
factor two at next-to-leading order (NLO). A lot of effort is put
in providing increasingly precise predictions. The first NLO
results were presented in the large top-quark mass limit more than
two-decades ago \cite{Dawson:1998py}. Full NLO QCD corrections including the
top-quark mass dependence were finally made available in \cite{Borowka:2016ehy,Borowka:2016ypz,Baglio:2018lrj,Baglio:2020ini}. The
next-to-next-to-leading order (NNLO) QCD corrections in the large $m_t$
limit were provided by
\cite{deFlorian:2013jea}, and the next-to-next-to-leading logarithmic
corrections in this limit by \cite{Shao:2013bz,deFlorian:2015moa}. 
Recently, the corrections due to the resummation of soft-gluon emission were provided up to next-to-next-to-next-to-leading logarithmic accuracy in \cite{Ajjath:2022kpv}.
The NNLO FT$_{\text{approx}}$\footnote{At
  FT$_{\text{approx}}$, the cross section is computed at
  next-to-next-to-leading order (NNLO) QCD in the heavy-top limit with
  full leading order (LO) and next-to-leading order (NLO) mass effects
  and full mass dependence in the one-loop double real corrections at NNLO QCD.} result was presented in 
\cite{Grazzini:2018bsd}. A combination of the usual renormalisation
and factorization scale uncertainties with the uncertainties
originating from the scheme and scale choice of the virtual top mass
was given in \cite{Baglio:2020wgt}. In the NMSSM we have additional
diagrams involving top and bottom squarks as well as the $s$-channel
exchange of non-SM-like Higgs bosons, {\it
  cf.}~Fig.~\ref{fig:doubleh}. In
\cite{Nhung:2013lpa,Abouabid:2021yvw} we computed the NLO  QCD
corrections in the heavy-top limit. \s

So far the complete electroweak (EW) corrections for gluon fusion into
Higgs pairs are not yet available. While in the
SM we can expect them to be of the 
order of a few percent by looking at the EW corrections to single
Higgs production
\cite{Djouadi:1994ge,Aglietti:2004nj,Degrassi:2004mx,Actis:2008ts,Actis:2008ug}
this might not be the case in beyond-the-SM models where couplings can be
enhanced compared to the SM or where light Higgs bosons could run in
the loops. The computation of the EW corrections to Higgs pair
production through gluon fusion is a major task and technical
challenge, which requires the computation of massive two-loop
integrals with several different mass scales. First steps have been
taken recently within the SM. In \cite{Muhlleitner:2022ijf} the top-Yukawa induced part
of the EW corrections and their relation to the effective trilinear
Higgs coupling have been provided and discussed. The subset of two-loop diagrams
where the Higgs boson is exchanged between the virtual top quark lines has been
calculated in the high-energy limit in \cite{Davies:2022ram}. \s

In this work we use our
effective loop-corrected Higgs self-couplings that make up part of
the EW corrections to get some insights on their importance. For this
we choose the parameter point {\tt BP10} where the di-Higgs cross section is
dominated by the resonant production of two SM-like Higgs bosons via
an intermediate heavy Higgs boson, as in this case the other diagrams
will give a subleading contribution to the total cross section so that the missing
EW higher-order corrections might be of less importance. In the
triangle diagram involving the resonant heavy Higgs boson in the
$s$-channel we still miss, however, the EW corrections to the top
triangle. \s

For this benchmark point we compute the gluon fusion production cross
section for a pair of SM-like Higgs bosons. We choose a c.m.~energy of
14~TeV, we use the CT14 pdf set \cite{Dulat:2015mca} and we set the top
quark mass to $m_t=172.74$~GeV. By using the
loop-corrected Higgs masses as inputs and the corresponding
higher-order corrected Higgs mixing angles to compute the Yukawa
couplings and the trilinear Higgs self-couplings that enter the
process through tree-level-like formulae,
we take into account the higher-order corrections to the input
parameters. By additionally including the loop-corrected trilinear
Higgs self-coupling computed in this paper we explicitly include
higher-order corrections to the observable, namely the Higgs pair
production process, though at an incomplete level as mentioned
above. \s

In Tab.~\ref{tab:cxnvalues} we compare the di-Higgs cross
sections for the case where only corrections to the input parameters
are considered (called 'inp' in the following) and the case where we
additionally include EW corrections to the process through the
loop-corrected trilinear Higgs self-couplings (called
'proc'). For simplicity we focus on the case where we include the full 1-loop
corrections both to the masses/mixing angles \cite{Graf:2012hh} and to the
trilinear Higgs self-couplings
\cite{Nhung:2013lpa,Muhlleitner:2015dua} (named '1L1L') 
and on the case where we include the 2-loop ${\cal O}(\alpha_t
(\alpha_s + \alpha_t))$ corrections both to the masses/mixing angles
\cite{Dao:2019qaz} and the trilinear Higgs self-couplings, computed in this
paper (named 'at2at2'). We furthermore list in the
table the values of the SM-like trilinear Higgs self-coupling,
$\lambda_{H_1 H_1 H_1}$, normalized to the SM-value for a Higgs boson
mass of same mass value, {\it i.e.}~$\lambda_{H_1 H_1
  H_1}^{\text{'SM'}}= 3 M_{H_1}^2/v$, called $\kappa_{H_1 H_1 H_1}$ in
the table. Accordingly, $\kappa_{H_2 H_1 H_1}$ is the $\lambda_{H_2
  H_1 H_1}$ coupling normalized to  $\lambda_{H_1 H_1
  H_1}^{\text{'SM'}}$. This value is also given in the table as the
resonant enhancement of the cross section basically comes from the
resonant $H_2$ production with subsequent decay into $H_1 H_1$. The
resonant production from the $H_3$ decay into $H_1 H_1$ plays only a
minor role for this benchmark point. We provide all these values both for OS and
$\overline{\mbox{DR}}$ renormalisation in the top/stop sector. Note
that in all renormalisation schemes and at all considered loop levels
the Yukawa coupling of the SM-like Higgs $H_1$ is practically SM-like
(it differs by only 1\% from the SM-value). \s
\begin{table}[t]
\begin{center}
\begin{tabular}{|l||c|c|c|c|c|c|c|}
\hline
'1L1L' & $\sigma^{\text{OS}}$ [fb] & $\sigma^{\overline{\text{DR}}}$  [fb] & $\kappa_{H_1 H_1
H_1}^{\text{OS}}$ & $\kappa_{H_1 H_1 H_1}^{\overline{\text{DR}}}$ & $\kappa_{H_2H_1 H_1}^{\text{OS}}$
  & $\kappa_{H_2 H_1 H_1}^{\overline{\text{DR}}}$ & $\Delta_{\text{ren}} \sigma$
   \\ \hline 
'inp' & 63.72 & 62.14 & 0.54 & 0.71 & -0.25 & -0.30 & 2.5\% \\ \hline
'proc' & 76.83 & 61.48 & 1.01 & 1.04 & -0.30 & -0.31 & 25\% \\ \hline \hline  
'at2at2'& $\sigma^{\text{OS}}$ [fb] & $\sigma^{\overline{\text{DR}}}$  [fb] & $\kappa_{H_1 H_1
H_1}^{\text{OS}}$ & $\kappa_{H_1 H_1 H_1}^{\overline{\text{DR}}}$ &
$\kappa_{H_2H_1 H_1}^{\text{OS}}$ & $\kappa_{H_2 H_1 H_1}^{\overline{\text{DR}}}$ & $\Delta_{\text{ren}} \sigma$ \\ \hline 
'inp' & 68.98 & 61.25 & 0.61 & 0.65 & -0.27 & -0.28 & 12.6\% \\ \hline
'proc' & 71.69 & 62.57 & 1.03 & 1.02 & -0.30 & -0.31 & 14.6\% \\ \hline
 \end{tabular}
\caption{{\tt BP10}: Cross section values for the production of a
  SM-like Higgs pair $H_1 H_1$ for OS and $\overline{\mbox{DR}}$
  renormalisation in the top/stop sector when using loop-corrected
  masses and mixing angles ('inp') and additionally loop-corrected
  effective trilinear Higgs self-couplings ('proc') at 1-loop order and at
  ${\cal O}(\alpha_t (\alpha_s + \alpha_t))$, respectively. The
  corresponding normalized trilinear Higgs self-couplings are given as
well as the relative change of
the cross section with the applied renormalisation scheme (see the text
for definition).} 
\label{tab:cxnvalues}
\end{center}
\end{table} 

From the cross section values we first of all notice the resonant
enhancement compared to the SM Higgs pair production cross section,
which at tree-level amounts to 19.72~fb. When we compare the absolute
values of the cross section we have to be careful as the changes not
only come from the use of different trilinear Higgs self-couplings and
renormalisation schemes, but also from the change in the kinematics as
the Higgs mass values depend on the loop order and the renormalisation
scheme. In the last column of Tab.~\ref{tab:cxnvalues} we give the
relative change of the cross section with respect to the applied
renormalisation scheme,
\beq
\Delta_{\text{ren}} \sigma \equiv
\frac{|\sigma^{m_t(\overline{\text{DR}})}-\sigma^{m_t(\text{OS})}|}{\sigma^{m_t(\overline{\text{DR}})}} \;.
\eeq
We observe that the inclusion of only the parameter corrections does
not decrease the renormalisation scheme dependence when moving from
one-loop order to two-loop ${\cal 
  O}(\alpha_t (\alpha_s+\alpha_t))$, on the contrary. This is not
astonishing as the scheme dependence of the input parameters has to be
compensated by the scheme dependence of the process-dependent
corrections at the same loop order. When these are included we observe
a decrease in the renormalisation scheme dependence of the cross
section at the same loop order when including higher and higher loop orders as
expected in perturbation theory. Still the renormalisation scheme
dependence with 14.6\% at ${\cal O}(\alpha_t (\alpha_s+\alpha_t))$ is
significant. This gives a hint that the remaining electroweak
corrections that we did not take into account in our approach might be
significant. It will hence be important to provide the complete EW
corrections to the cross section to be able to reduce the uncertainty
in its prediction due to missing higher loop corrections.

\section{Conclusions}
\label{sec:concl}
In this paper we presented the ${\cal O}(\alpha_t^2)$ corrections to
the trilinear Higgs self-couplings in the context of the CP-violating
NMSSM. They are part of our ongoing program of increasing the
precision in the predictions of the NMSSM Higgs potential parameters, the
masses and the Higgs self-couplings. We find that the corrections 
to the effective trilinear Higgs self-couplings are in general larger than those to the
Higgs boson masses. The relative corrections on top of the already
existing two-loop corrections at ${\cal O}(\alpha_t \alpha_s)$ are
much smaller, however, than when moving from one- to two-loop order
and indicate perturbative convergence. The remaining residual theoretical
uncertainties due to missing higher-order corrections, estimated from
the variation of the renormalisation scheme 
in the top/stop sector, range at the per-cent level and are also reduced
compared to the one-loop results. In general, the results obtained in
the $\DRbar$ scheme show a better convergence than in the OS scheme,
which is to be expected as they already partly resum higher-order
corrections. Within the theoretical uncertainties, the obtained
loop-corrected trilinear Higgs self-coupling of the SM-like Higgs
boson is in accordance with the result for the SM Higgs boson with same mass value.
The impact of the higher-order corrections on
the Higgs-to-Higgs decay widths is similar to the one on the
effective Higgs self-couplings. We also investigated the effect of the
inclusion of our loop-corrected effective Higgs self-couplings in the Higgs
pair production process. The estimates of the theoretical uncertainty
based on the variation of the renormalisation scheme indicate that the
remaining missing electroweak corrections to the process may be
significant. 

\section*{Acknowledgements}
The authors thank M.~Spira for discussions on higher-order corrections
to Higgs pair production.
M.M. and C.B. acknowledge support by the Deutsche
Forschungsgemeinschaft (DFG, German Research Foundation) under grant
396021762 - TRR 257. T.N.D.~is funded by the Vietnam
National Foundation for Science and Technology Development (NAFOSTED)
under grant number 103.01-2020.17. H.R.'s research is funded  by  the
Deutsche Forschungsgemeinschaft (DFG, German Research
Foundation)---project no.\ 442089526.
MG acknowledges support by the Deutsche Forschungsgemeinschaft
(DFG, German Research Foundation) under Germany’s Excellence
Strategy – EXC 2121 “Quantum Universe” – 390833306 and partially by 491245950.



\bibliographystyle{spphys_custom}
\bibliography{TrilNMSSM}

\begin{thebibliography}{100}
\providecommand{\url}[1]{{#1}}
\providecommand{\urlprefix}{URL }
\expandafter\ifx\csname urlstyle\endcsname\relax
  \providecommand{\doi}[1]{DOI \discretionary{}{}{}#1}\else
  \providecommand{\doi}{DOI \discretionary{}{}{}\begingroup
  \urlstyle{rm}\Url}\fi

\bibitem{DiMicco:2019ngk}
J.~Alison, et~al., \emph{{Higgs boson potential at colliders: Status and
  perspectives}}.
\newblock \href{https://doi.org/10.1016/j.revip.2020.100045}{Rev. Phys.
  \textbf{5}, 100045 (2020)},
  \href{https://arxiv.org/abs/1910.00012}{arXiv:1910.00012}

\bibitem{Djouadi:1999gv}
A.~Djouadi, W.~Kilian, M.~Muhlleitner, P.M. Zerwas, \emph{{Testing Higgs
  selfcouplings at e+ e- linear colliders}}.
\newblock \href{https://doi.org/10.1007/s100529900082}{Eur. Phys. J. C
  \textbf{10}, 27 (1999)},
  \href{https://arxiv.org/abs/hep-ph/9903229}{arXiv:hep-ph/9903229}

\bibitem{Djouadi:1999rca}
A.~Djouadi, W.~Kilian, M.~Muhlleitner, P.M. Zerwas, \emph{{Production of
  neutral Higgs boson pairs at LHC}}.
\newblock \href{https://doi.org/10.1007/s100529900083}{Eur. Phys. J. C
  \textbf{10}, 45 (1999)},
  \href{https://arxiv.org/abs/hep-ph/9904287}{arXiv:hep-ph/9904287}

\bibitem{Muhlleitner:2000jj}
M.M. Muhlleitner, \emph{{Higgs particles in the standard model and
  supersymmetric theories}}.
\newblock Ph.D. thesis, Hamburg U.,  2000.
\newblock \href{https://arxiv.org/abs/hep-ph/0008127}{arXiv:hep-ph/0008127}

\bibitem{Golfand:1971iw}
Y.~Golfand, E.~Likhtman, \emph{{Extension of the Algebra of Poincare Group
  Generators and Violation of p Invariance}}.
\newblock JETP Lett. \textbf{13}, 323 (1971)

\bibitem{Volkov:1973ix}
D.V. Volkov, V.P. Akulov, \emph{{Is the Neutrino a Goldstone Particle?}}
\newblock \href{https://doi.org/10.1016/0370-2693(73)90490-5}{Phys. Lett. B
  \textbf{46}, 109 (1973)}

\bibitem{Wess:1974tw}
J.~Wess, B.~Zumino, \emph{{Supergauge Transformations in Four-Dimensions}}.
\newblock \href{https://doi.org/10.1016/0550-3213(74)90355-1}{Nucl. Phys. B
  \textbf{70}, 39 (1974)}

\bibitem{Fayet:1974pd}
P.~Fayet, \emph{{Supergauge Invariant Extension of the Higgs Mechanism and a
  Model for the electron and Its Neutrino}}.
\newblock \href{https://doi.org/10.1016/0550-3213(75)90636-7}{Nucl. Phys. B
  \textbf{90}, 104 (1975)}

\bibitem{Fayet:1977yc}
P.~Fayet, \emph{{Spontaneously Broken Supersymmetric Theories of Weak,
  Electromagnetic and Strong Interactions}}.
\newblock \href{https://doi.org/10.1016/0370-2693(77)90852-8}{Phys. Lett. B
  \textbf{69}, 489 (1977)}

\bibitem{Fayet:1976cr}
P.~Fayet, S.~Ferrara, \emph{{Supersymmetry}}.
\newblock \href{https://doi.org/10.1016/0370-1573(77)90066-7}{Phys. Rept.
  \textbf{32}, 249 (1977)}

\bibitem{Nilles:1982dy}
H.P. Nilles, M.~Srednicki, D.~Wyler, \emph{{Weak Interaction Breakdown Induced
  by Supergravity}}.
\newblock \href{https://doi.org/10.1016/0370-2693(83)90460-4}{Phys. Lett. B
  \textbf{120}, 346 (1983)}

\bibitem{Nilles:1983ge}
H.P. Nilles, \emph{{Supersymmetry, Supergravity and Particle Physics}}.
\newblock \href{https://doi.org/10.1016/0370-1573(84)90008-5}{Phys. Rept.
  \textbf{110}, 1 (1984)}

\bibitem{Frere:1983ag}
J.M. Frere, D.R.T. Jones, S.~Raby, \emph{{Fermion Masses and Induction of the
  Weak Scale by Supergravity}}.
\newblock \href{https://doi.org/10.1016/0550-3213(83)90606-5}{Nucl. Phys. B
  \textbf{222}, 11 (1983)}

\bibitem{Derendinger:1983bz}
J.P. Derendinger, C.A. Savoy, \emph{{Quantum Effects and SU(2) x U(1) Breaking
  in Supergravity Gauge Theories}}.
\newblock \href{https://doi.org/10.1016/0550-3213(84)90162-7}{Nucl. Phys. B
  \textbf{237}, 307 (1984)}

\bibitem{Haber:1984rc}
H.E. Haber, G.L. Kane, \emph{{The Search for Supersymmetry: Probing Physics
  Beyond the Standard Model}}.
\newblock \href{https://doi.org/10.1016/0370-1573(85)90051-1}{Phys.Rept.
  \textbf{117}, 75 (1985)}

\bibitem{Sohnius:1985qm}
M.~Sohnius, \emph{{Introducing Supersymmetry}}.
\newblock \href{https://doi.org/10.1016/0370-1573(85)90023-7}{Phys.Rept.
  \textbf{128}, 39 (1985)}

\bibitem{Gunion:1984yn}
J.F. Gunion, H.E. Haber, \emph{{Higgs Bosons in Supersymmetric Models. 1.}}
\newblock \href{https://doi.org/10.1016/0550-3213(86)90340-8}{Nucl. Phys. B
  \textbf{272}, 1 (1986)} [Erratum: Nucl.Phys.B 402, 567--569 (1993)]

\bibitem{Gunion:1986nh}
J.F. Gunion, H.E. Haber, \emph{{Higgs Bosons in Supersymmetric Models. 2.
  Implications for Phenomenology}}.
\newblock \href{https://doi.org/10.1016/0550-3213(86)90050-7}{Nucl. Phys. B
  \textbf{278}, 449 (1986)} [Erratum: Nucl.Phys.B 402, 569--569 (1993)]

\bibitem{Gunion:1989we}
J.F. Gunion, H.E. Haber, G.L. Kane, S.~Dawson, \emph{{The Higgs Hunter's
  Guide}}.
\newblock Front. Phys. \textbf{80} (2000), 1--404

\bibitem{Martin:1997ns}
S.P. Martin, \emph{{A Supersymmetry primer}}.
\newblock \href{https://doi.org/10.1142/9789812839657_0001}{Adv. Ser. Direct.
  High Energy Phys. \textbf{18}, 1 (1998)},
  \href{https://arxiv.org/abs/hep-ph/9709356}{arXiv:hep-ph/9709356}

\bibitem{Dawson:1997tz}
S.~Dawson, \emph{{The MSSM and why it works}}, in \emph{{Theoretical Advanced
  Study Institute in Elementary Particle Physics (TASI 97): Supersymmetry,
  Supergravity and Supercolliders}} (1997), 261--339.
\newblock \href{https://arxiv.org/abs/hep-ph/9712464}{arXiv:hep-ph/9712464}

\bibitem{Djouadi:2005gj}
A.~Djouadi, \emph{{The Anatomy of electro-weak symmetry breaking. II. The Higgs
  bosons in the minimal supersymmetric model}}.
\newblock \href{https://doi.org/10.1016/j.physrep.2007.10.005}{Phys.Rept.
  \textbf{459}, 1 (2008)},
  \href{https://arxiv.org/abs/hep-ph/0503173}{arXiv:hep-ph/0503173}

\bibitem{HiggsAtlas}
G.~Aad, et~al., \emph{{Combined Measurement of the Higgs Boson Mass in $pp$
  Collisions at $\sqrt{s}=7$ and 8 TeV with the ATLAS and CMS Experiments}}.
\newblock \href{https://doi.org/10.1103/PhysRevLett.114.191803}{Phys. Rev.
  Lett. \textbf{114}, 191803 (2015)},
  \href{https://arxiv.org/abs/1503.07589}{arXiv:1503.07589}

\bibitem{Barbieri:1982eh}
R.~Barbieri, S.~Ferrara, C.A. Savoy, \emph{{Gauge Models with Spontaneously
  Broken Local Supersymmetry}}.
\newblock \href{https://doi.org/10.1016/0370-2693(82)90685-2}{Phys. Lett. B
  \textbf{119}, 343 (1982)}

\bibitem{Dine:1981rt}
M.~Dine, W.~Fischler, M.~Srednicki, \emph{{A Simple Solution to the Strong CP
  Problem with a Harmless Axion}}.
\newblock \href{https://doi.org/10.1016/0370-2693(81)90590-6}{Phys. Lett. B
  \textbf{104}, 199 (1981)}

\bibitem{Ellis:1988er}
J.R. Ellis, J.~Gunion, H.E. Haber, L.~Roszkowski, F.~Zwirner, \emph{{Higgs
  Bosons in a Nonminimal Supersymmetric Model}}.
\newblock \href{https://doi.org/10.1103/PhysRevD.39.844}{Phys. Rev. D
  \textbf{39}, 844 (1989)}

\bibitem{Drees:1988fc}
M.~Drees, \emph{{Supersymmetric Models with Extended Higgs Sector}}.
\newblock \href{https://doi.org/10.1142/S0217751X89001448}{Int. J. Mod. Phys. A
  \textbf{4}, 3635 (1989)}

\bibitem{Ellwanger:1993xa}
U.~Ellwanger, M.~Rausch~de Traubenberg, C.A. Savoy, \emph{{Particle spectrum in
  supersymmetric models with a gauge singlet}}.
\newblock \href{https://doi.org/10.1016/0370-2693(93)91621-S}{Phys. Lett. B
  \textbf{315}, 331 (1993)},
  \href{https://arxiv.org/abs/hep-ph/9307322}{arXiv:hep-ph/9307322}

\bibitem{Ellwanger:1995ru}
U.~Ellwanger, M.~Rausch~de Traubenberg, C.A. Savoy, \emph{{Higgs phenomenology
  of the supersymmetric model with a gauge singlet}}.
\newblock \href{https://doi.org/10.1007/BF01553993}{Z. Phys. C \textbf{67}, 665
  (1995)}, \href{https://arxiv.org/abs/hep-ph/9502206}{arXiv:hep-ph/9502206}

\bibitem{Ellwanger:1996gw}
U.~Ellwanger, M.~Rausch~de Traubenberg, C.A. Savoy, \emph{{Phenomenology of
  supersymmetric models with a singlet}}.
\newblock \href{https://doi.org/10.1016/S0550-3213(97)00128-4}{Nucl. Phys. B
  \textbf{492}, 21 (1997)},
  \href{https://arxiv.org/abs/hep-ph/9611251}{arXiv:hep-ph/9611251}

\bibitem{Elliott:1994ht}
T.~Elliott, S.F. King, P.L. White, \emph{{Unification constraints in the
  next-to-minimal supersymmetric standard model}}.
\newblock \href{https://doi.org/10.1016/0370-2693(95)00381-T}{Phys. Lett. B
  \textbf{351}, 213 (1995)},
  \href{https://arxiv.org/abs/hep-ph/9406303}{arXiv:hep-ph/9406303}

\bibitem{King:1995vk}
S.~King, P.~White, \emph{{Resolving the constrained minimal and next-to-minimal
  supersymmetric standard models}}.
\newblock \href{https://doi.org/10.1103/PhysRevD.52.4183}{Phys. Rev. D
  \textbf{52}, 4183 (1995)},
  \href{https://arxiv.org/abs/hep-ph/9505326}{arXiv:hep-ph/9505326}

\bibitem{Franke:1995tc}
F.~Franke, H.~Fraas, \emph{{Neutralinos and Higgs bosons in the next-to-minimal
  supersymmetric standard model}}.
\newblock \href{https://doi.org/10.1142/S0217751X97000529}{Int. J. Mod. Phys. A
  \textbf{12}, 479 (1997)},
  \href{https://arxiv.org/abs/hep-ph/9512366}{arXiv:hep-ph/9512366}

\bibitem{Maniatis:2009re}
M.~Maniatis, \emph{{The Next-to-Minimal Supersymmetric extension of the
  Standard Model reviewed}}.
\newblock \href{https://doi.org/10.1142/S0217751X10049827}{Int. J. Mod. Phys. A
  \textbf{25}, 3505 (2010)},
  \href{https://arxiv.org/abs/0906.0777}{arXiv:0906.0777}

\bibitem{Ellwanger:2009dp}
U.~Ellwanger, C.~Hugonie, A.M. Teixeira, \emph{{The Next-to-Minimal
  Supersymmetric Standard Model}}.
\newblock \href{https://doi.org/10.1016/j.physrep.2010.07.001}{Phys. Rept.
  \textbf{496}, 1 (2010)},
  \href{https://arxiv.org/abs/0910.1785}{arXiv:0910.1785}

\bibitem{Slavich:2020zjv}
P.~Slavich, et~al., \emph{{Higgs-mass predictions in the MSSM and beyond}}.
\newblock \href{https://doi.org/10.1140/epjc/s10052-021-09198-2}{Eur. Phys. J.
  C \textbf{81}(5), 450 (2021)},
  \href{https://arxiv.org/abs/2012.15629}{arXiv:2012.15629}

\bibitem{R:2021bml}
E.A.R. R., R.~Fazio, \emph{{High-Precision Calculations of the Higgs Boson
  Mass}}.
\newblock \href{https://doi.org/10.3390/particles5010006}{Particles
  \textbf{5}(1), 53 (2022)},
  \href{https://arxiv.org/abs/2112.15295}{arXiv:2112.15295}

\bibitem{Barger:1991ed}
V.D. Barger, M.S. Berger, A.L. Stange, R.J.N. Phillips, \emph{{Supersymmetric
  Higgs boson hadroproduction and decays including radiative corrections}}.
\newblock \href{https://doi.org/10.1103/PhysRevD.45.4128}{Phys. Rev. D
  \textbf{45}, 4128 (1992)}

\bibitem{Hollik:2001px}
W.~Hollik, S.~Penaranda, \emph{{Yukawa coupling quantum corrections to the
  selfcouplings of the lightest MSSM Higgs boson}}.
\newblock \href{https://doi.org/10.1007/s100520100862}{Eur. Phys. J. C
  \textbf{23}, 163 (2002)},
  \href{https://arxiv.org/abs/hep-ph/0108245}{arXiv:hep-ph/0108245}

\bibitem{Dobado:2002jz}
A.~Dobado, M.J. Herrero, W.~Hollik, S.~Penaranda, \emph{{Selfinteractions of
  the lightest MSSM Higgs boson in the large pseudoscalar mass limit}}.
\newblock \href{https://doi.org/10.1103/PhysRevD.66.095016}{Phys. Rev. D
  \textbf{66}, 095016 (2002)},
  \href{https://arxiv.org/abs/hep-ph/0208014}{arXiv:hep-ph/0208014}

\bibitem{Williams:2007dc}
K.E. Williams, G.~Weiglein, \emph{{Precise predictions for $h_{a} \to h_{b}
  h_{c}$ decays in the complex MSSM}}.
\newblock \href{https://doi.org/10.1016/j.physletb.2007.12.049}{Phys. Lett. B
  \textbf{660}, 217 (2008)},
  \href{https://arxiv.org/abs/0710.5320}{arXiv:0710.5320}

\bibitem{Williams:2011bu}
K.E. Williams, H.~Rzehak, G.~Weiglein, \emph{{Higher order corrections to Higgs
  boson decays in the MSSM with complex parameters}}.
\newblock \href{https://doi.org/10.1140/epjc/s10052-011-1669-3}{Eur. Phys. J. C
  \textbf{71}, 1669 (2011)},
  \href{https://arxiv.org/abs/1103.1335}{arXiv:1103.1335}

\bibitem{Brucherseifer:2013qva}
M.~Brucherseifer, R.~Gavin, M.~Spira, \emph{{Minimal supersymmetric Higgs boson
  self-couplings: Two-loop $O(\alpha_{t}\alpha_{s})$ corrections}}.
\newblock \href{https://doi.org/10.1103/PhysRevD.90.117701}{Phys. Rev. D
  \textbf{90}(11), 117701 (2014)},
  \href{https://arxiv.org/abs/1309.3140}{arXiv:1309.3140}

\bibitem{Nhung:2013lpa}
D.T. Nhung, M.~Muhlleitner, J.~Streicher, K.~Walz, \emph{{Higher Order
  Corrections to the Trilinear Higgs Self-Couplings in the Real NMSSM}}.
\newblock \href{https://doi.org/10.1007/JHEP11(2013)181}{JHEP \textbf{1311},
  181 (2013)}, \href{https://arxiv.org/abs/1306.3926}{arXiv:1306.3926}

\bibitem{Muhlleitner:2015dua}
M.~M\"uhlleitner, D.T. Nhung, H.~Ziesche, \emph{{The order $
  \mathcal{O}\left({\alpha}_t{\alpha}_s\right) $ corrections to the trilinear
  Higgs self-couplings in the complex NMSSM}}.
\newblock \href{https://doi.org/10.1007/JHEP12(2015)034}{JHEP \textbf{12}, 034
  (2015)}, \href{https://arxiv.org/abs/1506.03321}{arXiv:1506.03321}

\bibitem{Baglio:2019nlc}
J.~Baglio, T.N. Dao, M.~M\"uhlleitner, \emph{{One-Loop Corrections to the
  Two-Body Decays of the Neutral Higgs Bosons in the Complex NMSSM}}.
\newblock \href{https://doi.org/10.1140/epjc/s10052-020-08520-8}{Eur. Phys. J.
  C \textbf{80}(10), 960 (2020)},
  \href{https://arxiv.org/abs/1907.12060}{arXiv:1907.12060}

\bibitem{Kanemura:2002vm}
S.~Kanemura, S.~Kiyoura, Y.~Okada, E.~Senaha, C.P. Yuan, \emph{{New physics
  effect on the Higgs selfcoupling}}.
\newblock \href{https://doi.org/10.1016/S0370-2693(03)00268-5}{Phys. Lett. B
  \textbf{558}, 157 (2003)},
  \href{https://arxiv.org/abs/hep-ph/0211308}{arXiv:hep-ph/0211308}

\bibitem{Kanemura:2004mg}
S.~Kanemura, Y.~Okada, E.~Senaha, C.P. Yuan, \emph{{Higgs coupling constants as
  a probe of new physics}}.
\newblock \href{https://doi.org/10.1103/PhysRevD.70.115002}{Phys. Rev. D
  \textbf{70}, 115002 (2004)},
  \href{https://arxiv.org/abs/hep-ph/0408364}{arXiv:hep-ph/0408364}

\bibitem{Kanemura:2015mxa}
S.~Kanemura, M.~Kikuchi, K.~Yagyu, \emph{{Fingerprinting the extended Higgs
  sector using one-loop corrected Higgs boson couplings and future precision
  measurements}}.
\newblock \href{https://doi.org/10.1016/j.nuclphysb.2015.04.015}{Nucl. Phys. B
  \textbf{896}, 80 (2015)},
  \href{https://arxiv.org/abs/1502.07716}{arXiv:1502.07716}

\bibitem{Kanemura:2017wtm}
S.~Kanemura, M.~Kikuchi, K.~Sakurai, K.~Yagyu, \emph{{Gauge invariant one-loop
  corrections to Higgs boson couplings in non-minimal Higgs models}}.
\newblock \href{https://doi.org/10.1103/PhysRevD.96.035014}{Phys. Rev. D
  \textbf{96}(3), 035014 (2017)},
  \href{https://arxiv.org/abs/1705.05399}{arXiv:1705.05399}

\bibitem{Basler:2017uxn}
P.~Basler, M.~M\"uhlleitner, J.~Wittbrodt, \emph{{The CP-Violating 2HDM in
  Light of a Strong First Order Electroweak Phase Transition and Implications
  for Higgs Pair Production}}.
\newblock \href{https://doi.org/10.1007/JHEP03(2018)061}{JHEP \textbf{03}, 061
  (2018)}, \href{https://arxiv.org/abs/1711.04097}{arXiv:1711.04097}

\bibitem{Basler:2019iuu}
P.~Basler, M.~M\"uhlleitner, J.~M\"uller, \emph{{Electroweak Phase Transition
  in Non-Minimal Higgs Sectors}}.
\newblock \href{https://doi.org/10.1007/JHEP05(2020)016}{JHEP \textbf{05}, 016
  (2020)}, \href{https://arxiv.org/abs/1912.10477}{arXiv:1912.10477}

\bibitem{Senaha:2018xek}
E.~Senaha, \emph{{Radiative Corrections to Triple Higgs Coupling and
  Electroweak Phase Transition: Beyond One-loop Analysis}}.
\newblock \href{https://doi.org/10.1103/PhysRevD.100.055034}{Phys. Rev. D
  \textbf{100}(5), 055034 (2019)},
  \href{https://arxiv.org/abs/1811.00336}{arXiv:1811.00336}

\bibitem{Braathen:2019pxr}
J.~Braathen, S.~Kanemura, \emph{{On two-loop corrections to the Higgs trilinear
  coupling in models with extended scalar sectors}}.
\newblock \href{https://doi.org/10.1016/j.physletb.2019.07.021}{Phys. Lett. B
  \textbf{796}, 38 (2019)},
  \href{https://arxiv.org/abs/1903.05417}{arXiv:1903.05417}

\bibitem{Braathen:2019zoh}
J.~Braathen, S.~Kanemura, \emph{{Leading two-loop corrections to the Higgs
  boson self-couplings in models with extended scalar sectors}}.
\newblock \href{https://doi.org/10.1140/epjc/s10052-020-7723-2}{Eur. Phys. J. C
  \textbf{80}(3), 227 (2020)},
  \href{https://arxiv.org/abs/1911.11507}{arXiv:1911.11507}

\bibitem{Bahl:2022jnx}
H.~Bahl, J.~Braathen, G.~Weiglein, \emph{{New constraints on extended Higgs
  sectors from the trilinear Higgs coupling}}.
\newblock \href{https://arxiv.org/abs/2202.03453}{arXiv:2202.03453}

\bibitem{Krause:2016oke}
M.~Krause, R.~Lorenz, M.~Muhlleitner, R.~Santos, H.~Ziesche,
  \emph{{Gauge-independent Renormalization of the 2-Higgs-Doublet Model}}.
\newblock \href{https://doi.org/10.1007/JHEP09(2016)143}{JHEP \textbf{09}, 143
  (2016)}, \href{https://arxiv.org/abs/1605.04853}{arXiv:1605.04853}

\bibitem{Bojarski:2015kra}
F.~Bojarski, G.~Chalons, D.~Lopez-Val, T.~Robens, \emph{{Heavy to light Higgs
  boson decays at NLO in the Singlet Extension of the Standard Model}}.
\newblock \href{https://doi.org/10.1007/JHEP02(2016)147}{JHEP \textbf{02}, 147
  (2016)}, \href{https://arxiv.org/abs/1511.08120}{arXiv:1511.08120}

\bibitem{Krause:2017mal}
M.~Krause, D.~Lopez-Val, M.~Muhlleitner, R.~Santos, \emph{{Gauge-independent
  Renormalization of the N2HDM}}.
\newblock \href{https://doi.org/10.1007/JHEP12(2017)077}{JHEP \textbf{12}, 077
  (2017)}, \href{https://arxiv.org/abs/1708.01578}{arXiv:1708.01578}

\bibitem{Krause:2018wmo}
M.~Krause, M.~M\"uhlleitner, M.~Spira, \emph{{2HDECAY \textemdash{}A program
  for the calculation of electroweak one-loop corrections to Higgs decays in
  the Two-Higgs-Doublet Model including state-of-the-art QCD corrections}}.
\newblock \href{https://doi.org/10.1016/j.cpc.2019.08.003}{Comput. Phys.
  Commun. \textbf{246}, 106852 (2020)},
  \href{https://arxiv.org/abs/1810.00768}{arXiv:1810.00768}

\bibitem{Denner:2018opp}
A.~Denner, S.~Dittmaier, J.N. Lang, \emph{{Renormalization of mixing angles}}.
\newblock \href{https://doi.org/10.1007/JHEP11(2018)104}{JHEP \textbf{11}, 104
  (2018)}, \href{https://arxiv.org/abs/1808.03466}{arXiv:1808.03466}

\bibitem{Krause:2019oar}
M.~Krause, M.~M\"uhlleitner, \emph{{ewN2HDECAY - A program for the Calculation
  of Electroweak One-Loop Corrections to Higgs Decays in the Next-to-Minimal
  Two-Higgs-Doublet Model Including State-of-the-Art QCD Corrections}}.
\newblock \href{https://doi.org/10.1016/j.cpc.2019.106924}{Comput. Phys.
  Commun. \textbf{247}, 106924 (2020)},
  \href{https://arxiv.org/abs/1904.02103}{arXiv:1904.02103}

\bibitem{Krause:2019qwe}
M.~Krause, M.~M\"uhlleitner, \emph{{Impact of Electroweak Corrections on
  Neutral Higgs Boson Decays in Extended Higgs Sectors}}.
\newblock \href{https://doi.org/10.1007/JHEP04(2020)083}{JHEP \textbf{04}, 083
  (2020)}, \href{https://arxiv.org/abs/1912.03948}{arXiv:1912.03948}

\bibitem{Azevedo:2021ylf}
D.~Azevedo, P.~Gabriel, M.~Muhlleitner, K.~Sakurai, R.~Santos, \emph{{One-loop
  corrections to the Higgs boson invisible decay in the dark doublet phase of
  the N2HDM}}.
\newblock \href{https://doi.org/10.1007/JHEP10(2021)044}{JHEP \textbf{10}, 044
  (2021)}, \href{https://arxiv.org/abs/2104.03184}{arXiv:2104.03184}

\bibitem{Egle:2022wmq}
F.~Egle, M.~M\"uhlleitner, R.~Santos, J.a. Viana, \emph{{One-loop Corrections
  to the Higgs Boson Invisible Decay in a Complex Singlet Extension of the
  SM}}.
\newblock \href{https://arxiv.org/abs/2202.04035}{arXiv:2202.04035}

\bibitem{Goodsell:2017pdq}
M.D. Goodsell, S.~Liebler, F.~Staub, \emph{{Generic calculation of two-body
  partial decay widths at the full one-loop level}}.
\newblock \href{https://doi.org/10.1140/epjc/s10052-017-5259-x}{Eur. Phys. J. C
  \textbf{77}(11), 758 (2017)},
  \href{https://arxiv.org/abs/1703.09237}{arXiv:1703.09237}

\bibitem{Baglio:2013iia}
J.~Baglio, R.~Gr\"ober, M.~M\"uhlleitner, D.T. Nhung, H.~Rzehak, M.~Spira,
  J.~Streicher, K.~Walz, \emph{{NMSSMCALC: A Program Package for the
  Calculation of Loop-Corrected Higgs Boson Masses and Decay Widths in the
  (Complex) NMSSM}}.
\newblock \href{https://doi.org/10.1016/j.cpc.2014.08.005}{Comput. Phys.
  Commun. \textbf{185}(12), 3372 (2014)},
  \href{https://arxiv.org/abs/1312.4788}{arXiv:1312.4788}

\bibitem{King:2015oxa}
S.F. King, M.~Muhlleitner, R.~Nevzorov, K.~Walz, \emph{{Exploring the
  CP-violating NMSSM: EDM Constraints and Phenomenology}}.
\newblock \href{https://doi.org/10.1016/j.nuclphysb.2015.11.003}{Nucl. Phys. B
  \textbf{901}, 526 (2015)},
  \href{https://arxiv.org/abs/1508.03255}{arXiv:1508.03255}

\bibitem{Dao:2019qaz}
T.~Dao, R.~Gröber, M.~Krause, M.~Mühlleitner, H.~Rzehak, \emph{{Two-loop $
  \mathcal{O} $ ( $ {\alpha}_t^2 $ ) corrections to the neutral Higgs boson
  masses in the CP-violating NMSSM}}.
\newblock \href{https://doi.org/10.1007/JHEP08(2019)114}{JHEP \textbf{08}, 114
  (2019)}

\bibitem{Muhlleitner:2014vsa}
M.~Mühlleitner, D.T. Nhung, H.~Rzehak, K.~Walz, \emph{{Two-loop contributions
  of the order $ \mathcal{O}\left({\alpha}_t{\alpha}_s\right) $ to the masses
  of the Higgs bosons in the CP-violating NMSSM}}.
\newblock \href{https://doi.org/10.1007/JHEP05(2015)128}{JHEP \textbf{05}, 128
  (2015)}, \href{https://arxiv.org/abs/1412.0918}{arXiv:1412.0918}

\bibitem{Dao:2021khm}
T.N. Dao, M.~Gabelmann, M.~M\"uhlleitner, H.~Rzehak, \emph{{Two-loop $
  \mathcal{O} $((\ensuremath{\alpha}$_{t}$ + \ensuremath{\alpha}$_{\lambda}$ +
  \ensuremath{\alpha}$_{\kappa}$)$^{2}$) corrections to the Higgs boson masses
  in the CP-violating NMSSM}}.
\newblock \href{https://doi.org/10.1007/JHEP09(2021)193}{JHEP \textbf{09}, 193
  (2021)}, \href{https://arxiv.org/abs/2106.06990}{arXiv:2106.06990}

\bibitem{Skands:2003cj}
P.Z. Skands, et~al., \emph{{SUSY Les Houches accord: Interfacing SUSY spectrum
  calculators, decay packages, and event generators}}.
\newblock \href{https://doi.org/10.1088/1126-6708/2004/07/036}{JHEP
  \textbf{07}, 036 (2004)}

\bibitem{Allanach:2008qq}
B.C. Allanach, et~al., \emph{{SUSY Les Houches Accord 2}}.
\newblock \href{https://doi.org/10.1016/j.cpc.2008.08.004}{Comput. Phys.
  Commun. \textbf{180}, 8 (2009)},
  \href{https://arxiv.org/abs/0801.0045}{arXiv:0801.0045}

\bibitem{Ender:2011qh}
K.~Ender, T.~Graf, M.~Muhlleitner, H.~Rzehak, \emph{{Analysis of the NMSSM
  Higgs Boson Masses at One-Loop Level}}.
\newblock \href{https://doi.org/10.1103/PhysRevD.85.075024}{Phys. Rev. D
  \textbf{85}, 075024 (2012)}

\bibitem{Graf:2012hh}
T.~Graf, R.~Grober, M.~Muhlleitner, H.~Rzehak, K.~Walz, \emph{{Higgs Boson
  Masses in the Complex NMSSM at One-Loop Level}}.
\newblock \href{https://doi.org/10.1007/JHEP10(2012)122}{JHEP \textbf{10}, 122
  (2012)}

\bibitem{Davydychev:1992mt}
A.I. Davydychev, J.~Tausk, \emph{{Two loop selfenergy diagrams with different
  masses and the momentum expansion}}.
\newblock \href{https://doi.org/10.1016/0550-3213(93)90338-P}{Nucl. Phys. B
  \textbf{397}, 123 (1993)}

\bibitem{Ford:1992pn}
C.~Ford, I.~Jack, D.~Jones, \emph{{The Standard model effective potential at
  two loops}}.
\newblock \href{https://doi.org/10.1016/0550-3213(92)90165-8}{Nucl. Phys. B
  \textbf{387}, 373 (1992)} [Erratum: Nucl.Phys.B 504, 551--552 (1997)]

\bibitem{Scharf:1993ds}
R.~Scharf, J.~Tausk, \emph{{Scalar two loop integrals for gauge boson
  selfenergy diagrams with a massless fermion loop}}.
\newblock \href{https://doi.org/10.1016/0550-3213(94)90391-3}{Nucl. Phys. B
  \textbf{412}, 523 (1994)}

\bibitem{Weiglein:1993hd}
G.~Weiglein, R.~Scharf, M.~Bohm, \emph{{Reduction of general two loop
  selfenergies to standard scalar integrals}}.
\newblock \href{https://doi.org/10.1016/0550-3213(94)90325-5}{Nucl. Phys. B
  \textbf{416}, 606 (1994)},
  \href{https://arxiv.org/abs/hep-ph/9310358}{arXiv:hep-ph/9310358}

\bibitem{Berends:1994ed}
F.A. Berends, J.B. Tausk, \emph{{On the numerical evaluation of scalar two loop
  selfenergy diagrams}}.
\newblock \href{https://doi.org/10.1016/0550-3213(94)90336-0}{Nucl. Phys. B
  \textbf{421}, 456 (1994)}

\bibitem{Martin:2001vx}
S.P. Martin, \emph{{Two Loop Effective Potential for a General Renormalizable
  Theory and Softly Broken Supersymmetry}}.
\newblock \href{https://doi.org/10.1103/PhysRevD.65.116003}{Phys. Rev. D
  \textbf{65}, 116003 (2002)},
  \href{https://arxiv.org/abs/hep-ph/0111209}{arXiv:hep-ph/0111209}

\bibitem{Martin:2005qm}
S.P. Martin, D.G. Robertson, \emph{{TSIL: A Program for the calculation of
  two-loop self-energy integrals}}.
\newblock \href{https://doi.org/10.1016/j.cpc.2005.08.005}{Comput. Phys.
  Commun. \textbf{174}, 133 (2006)}

\bibitem{Staub:2008uz}
F.~Staub, \emph{{SARAH}}.
\newblock \href{https://arxiv.org/abs/0806.0538}{arXiv:0806.0538}

\bibitem{Staub:2010jh}
F.~Staub, \emph{{Automatic Calculation of supersymmetric Renormalization Group
  Equations and Self Energies}}.
\newblock \href{https://doi.org/10.1016/j.cpc.2010.11.030}{Comput. Phys.
  Commun. \textbf{182}, 808 (2011)},
  \href{https://arxiv.org/abs/1002.0840}{arXiv:1002.0840}

\bibitem{Staub:2012pb}
F.~Staub, \emph{{SARAH 3.2: Dirac Gauginos, UFO output, and more}}.
\newblock \href{https://doi.org/10.1016/j.cpc.2013.02.019}{Computer Physics
  Communications \textbf{184}, pp. 1792 (2013)},
  \href{https://arxiv.org/abs/1207.0906}{arXiv:1207.0906}

\bibitem{Staub:2013tta}
F.~Staub, \emph{{SARAH 4 : A tool for (not only SUSY) model builders}}.
\newblock \href{https://doi.org/10.1016/j.cpc.2014.02.018}{Comput. Phys.
  Commun. \textbf{185}, 1773 (2014)},
  \href{https://arxiv.org/abs/1309.7223}{arXiv:1309.7223}

\bibitem{Goodsell:2014bna}
M.D. Goodsell, K.~Nickel, F.~Staub, \emph{{Two-Loop Higgs mass calculations in
  supersymmetric models beyond the MSSM with SARAH and SPheno}}.
\newblock \href{https://doi.org/10.1140/epjc/s10052-014-3247-y}{Eur. Phys. J. C
  \textbf{75}(1), 32 (2015)},
  \href{https://arxiv.org/abs/1411.0675}{arXiv:1411.0675}

\bibitem{Goodsell:2014pla}
M.D. Goodsell, K.~Nickel, F.~Staub, \emph{{Two-loop corrections to the Higgs
  masses in the NMSSM}}.
\newblock \href{https://doi.org/10.1103/PhysRevD.91.035021}{Phys. Rev. D
  \textbf{91}, 035021 (2015)},
  \href{https://arxiv.org/abs/1411.4665}{arXiv:1411.4665}

\bibitem{Kublbeck:1990xc}
J.~Kublbeck, M.~Bohm, A.~Denner, \emph{{Feyn Arts: Computer Algebraic
  Generation of Feynman Graphs and Amplitudes}}.
\newblock
  \href{https://doi.org/10.1016/0010-4655(90)90001-H}{Comput.Phys.Commun.
  \textbf{60}, 165 (1990)}

\bibitem{Hahn:2000kx}
T.~Hahn, \emph{{Generating Feynman diagrams and amplitudes with FeynArts 3}}.
\newblock
  \href{https://doi.org/10.1016/S0010-4655(01)00290-9}{Comput.Phys.Commun.
  \textbf{140}, 418 (2001)},
  \href{https://arxiv.org/abs/hep-ph/0012260}{arXiv:hep-ph/0012260}

\bibitem{Mertig:1990an}
R.~Mertig, M.~Bohm, A.~Denner, \emph{{FEYN CALC: Computer algebraic calculation
  of Feynman amplitudes}}.
\newblock
  \href{https://doi.org/10.1016/0010-4655(91)90130-D}{Comput.Phys.Commun.
  \textbf{64}, 345 (1991)}

\bibitem{Shtabovenko:2016sxi}
V.~Shtabovenko, R.~Mertig, F.~Orellana, \emph{{New Developments in FeynCalc
  9.0}}.
\newblock \href{https://doi.org/10.1016/j.cpc.2016.06.008}{Comput. Phys.
  Commun. \textbf{207}, 432 (2016)},
  \href{https://arxiv.org/abs/1601.01167}{arXiv:1601.01167}

\bibitem{Mertig:1998vk}
R.~Mertig, R.~Scharf, \emph{{TARCER: A Mathematica program for the reduction of
  two loop propagator integrals}}.
\newblock
  \href{https://doi.org/10.1016/S0010-4655(98)00042-3}{Comput.Phys.Commun.
  \textbf{111}, 265 (1998)},
  \href{https://arxiv.org/abs/hep-ph/9801383}{arXiv:hep-ph/9801383}

\bibitem{Bechtle:2008jh}
P.~Bechtle, O.~Brein, S.~Heinemeyer, G.~Weiglein, K.E. Williams,
  \emph{{HiggsBounds: Confronting Arbitrary Higgs Sectors with Exclusion Bounds
  from LEP and the Tevatron}}.
\newblock \href{https://doi.org/10.1016/j.cpc.2009.09.003}{Comput. Phys.
  Commun. \textbf{181}, 138 (2010)},
  \href{https://arxiv.org/abs/0811.4169}{arXiv:0811.4169}

\bibitem{Bechtle:2011sb}
P.~Bechtle, O.~Brein, S.~Heinemeyer, G.~Weiglein, K.E. Williams,
  \emph{{HiggsBounds 2.0.0: Confronting Neutral and Charged Higgs Sector
  Predictions with Exclusion Bounds from LEP and the Tevatron}}.
\newblock \href{https://doi.org/10.1016/j.cpc.2011.07.015}{Comput. Phys.
  Commun. \textbf{182}, 2605 (2011)},
  \href{https://arxiv.org/abs/1102.1898}{arXiv:1102.1898}

\bibitem{Bechtle:2013wla}
P.~Bechtle, O.~Brein, S.~Heinemeyer, O.~St\r{a}l, T.~Stefaniak, G.~Weiglein,
  K.E. Williams, \emph{{$\mathsf{HiggsBounds}-4$: Improved Tests of Extended
  Higgs Sectors against Exclusion Bounds from LEP, the Tevatron and the LHC}}.
\newblock \href{https://doi.org/10.1140/epjc/s10052-013-2693-2}{Eur. Phys. J. C
  \textbf{74}(3), 2693 (2014)},
  \href{https://arxiv.org/abs/1311.0055}{arXiv:1311.0055}

\bibitem{Bechtle:2013xfa}
P.~Bechtle, S.~Heinemeyer, O.~St\r{a}l, T.~Stefaniak, G.~Weiglein,
  \emph{{$HiggsSignals$: Confronting arbitrary Higgs sectors with measurements
  at the Tevatron and the LHC}}.
\newblock \href{https://doi.org/10.1140/epjc/s10052-013-2711-4}{Eur. Phys. J. C
  \textbf{74}(2), 2711 (2014)},
  \href{https://arxiv.org/abs/1305.1933}{arXiv:1305.1933}

\bibitem{PhysRevD.98.030001}
M.~Tanabashi, et~al., \emph{Review of particle physics}.
\newblock \href{https://doi.org/10.1103/PhysRevD.98.030001}{Phys. Rev. D
  \textbf{98}, 030001 (2018)}

\bibitem{Dennerlhcnote}
A.~Denner, S.~Dittmaier, M.~Grazzini, R.V. Harlander, R.S. Thorne, M.~Spira,
  M.~Steinhauser, \emph{{Standard Model input parameters for Higgs physics}}.
\newblock \href{https://cds.cern.ch/record/2047636}{LHCHXSWG-INT-2015-006}

\bibitem{Abouabid:2021yvw}
H.~Abouabid, A.~Arhrib, D.~Azevedo, J.E. Falaki, P.M. Ferreira,
  M.~M\"uhlleitner, R.~Santos, \emph{{Benchmarking Di-Higgs Production in
  Various Extended Higgs Sector Models}}.
\newblock \href{https://arxiv.org/abs/2112.12515}{arXiv:2112.12515}

\bibitem{atlaspaperdihiggs}
A.~Collaboration, \emph{{Constraining the Higgs boson self-coupling from
  single- and double-Higgs production with the ATLAS detector using pp
  collisions at sqrt(s)=13 TeV}}.
\newblock \href{http://cds.cern.ch/record/2816332}{ATLAS-CONF-2022-50}

\bibitem{CMS:2022hgz}
C.~Collaboration, \emph{{Search for nonresonant Higgs boson pair production in
  final state with two bottom quarks and two tau leptons in proton-proton
  collisions at $\sqrt{s}$ = 13 TeV}}.
\newblock \href{https://arxiv.org/abs/2206.09401}{arXiv:2206.09401}

\bibitem{deFlorian:2016spz}
D.~de~Florian, et~al., \emph{{Handbook of LHC Higgs Cross Sections: 4.
  Deciphering the Nature of the Higgs Sector}}.
\newblock \href{https://doi.org/10.23731/CYRM-2017-002}{CERN Yellow Reports:
  Monographs \textbf{2/2017}},
  \href{https://arxiv.org/abs/1610.07922}{arXiv:1610.07922}

\bibitem{Baglio:2012np}
J.~Baglio, A.~Djouadi, R.~Gr\"ober, M.M. M\"uhlleitner, J.~Quevillon, M.~Spira,
  \emph{{The measurement of the Higgs self-coupling at the LHC: theoretical
  status}}.
\newblock \href{https://doi.org/10.1007/JHEP04(2013)151}{JHEP \textbf{04}, 151
  (2013)}, \href{https://arxiv.org/abs/1212.5581}{arXiv:1212.5581}

\bibitem{Dawson:1998py}
S.~Dawson, S.~Dittmaier, M.~Spira, \emph{{Neutral Higgs boson pair production
  at hadron colliders: QCD corrections}}.
\newblock \href{https://doi.org/10.1103/PhysRevD.58.115012}{Phys. Rev. D
  \textbf{58}, 115012 (1998)},
  \href{https://arxiv.org/abs/hep-ph/9805244}{arXiv:hep-ph/9805244}

\bibitem{Borowka:2016ehy}
S.~Borowka, N.~Greiner, G.~Heinrich, S.P. Jones, M.~Kerner, J.~Schlenk,
  U.~Schubert, T.~Zirke, \emph{{Higgs Boson Pair Production in Gluon Fusion at
  Next-to-Leading Order with Full Top-Quark Mass Dependence}}.
\newblock \href{https://doi.org/10.1103/PhysRevLett.117.079901}{Phys. Rev.
  Lett. \textbf{117}(1), 012001 (2016)} [Erratum: Phys.Rev.Lett. 117, 079901
  (2016)], \href{https://arxiv.org/abs/1604.06447}{arXiv:1604.06447}

\bibitem{Borowka:2016ypz}
S.~Borowka, N.~Greiner, G.~Heinrich, S.P. Jones, M.~Kerner, J.~Schlenk,
  T.~Zirke, \emph{{Full top quark mass dependence in Higgs boson pair
  production at NLO}}.
\newblock \href{https://doi.org/10.1007/JHEP10(2016)107}{JHEP \textbf{10}, 107
  (2016)}, \href{https://arxiv.org/abs/1608.04798}{arXiv:1608.04798}

\bibitem{Baglio:2018lrj}
J.~Baglio, F.~Campanario, S.~Glaus, M.~M\"uhlleitner, M.~Spira, J.~Streicher,
  \emph{{Gluon fusion into Higgs pairs at NLO QCD and the top mass scheme}}.
\newblock \href{https://doi.org/10.1140/epjc/s10052-019-6973-3}{Eur. Phys. J. C
  \textbf{79}(6), 459 (2019)},
  \href{https://arxiv.org/abs/1811.05692}{arXiv:1811.05692}

\bibitem{Baglio:2020ini}
J.~Baglio, F.~Campanario, S.~Glaus, M.~M\"uhlleitner, J.~Ronca, M.~Spira,
  J.~Streicher, \emph{{Higgs-Pair Production via Gluon Fusion at Hadron
  Colliders: NLO QCD Corrections}}.
\newblock \href{https://doi.org/10.1007/JHEP04(2020)181}{JHEP \textbf{04}, 181
  (2020)}, \href{https://arxiv.org/abs/2003.03227}{arXiv:2003.03227}

\bibitem{deFlorian:2013jea}
D.~de~Florian, J.~Mazzitelli, \emph{{Higgs Boson Pair Production at
  Next-to-Next-to-Leading Order in QCD}}.
\newblock \href{https://doi.org/10.1103/PhysRevLett.111.201801}{Phys. Rev.
  Lett. \textbf{111}, 201801 (2013)},
  \href{https://arxiv.org/abs/1309.6594}{arXiv:1309.6594}

\bibitem{Shao:2013bz}
D.Y. Shao, C.S. Li, H.T. Li, J.~Wang, \emph{{Threshold resummation effects in
  Higgs boson pair production at the LHC}}.
\newblock \href{https://doi.org/10.1007/JHEP07(2013)169}{JHEP \textbf{07}, 169
  (2013)}, \href{https://arxiv.org/abs/1301.1245}{arXiv:1301.1245}

\bibitem{deFlorian:2015moa}
D.~de~Florian, J.~Mazzitelli, \emph{{Higgs pair production at
  next-to-next-to-leading logarithmic accuracy at the LHC}}.
\newblock \href{https://doi.org/10.1007/JHEP09(2015)053}{JHEP \textbf{09}, 053
  (2015)}, \href{https://arxiv.org/abs/1505.07122}{arXiv:1505.07122}

\bibitem{Ajjath:2022kpv}
A.H. Ajjath, H.S. Shao, \emph{{N$^3$LO+N$^3$LL QCD improved Higgs pair cross
  sections}}.
\newblock \href{https://arxiv.org/abs/2209.03914}{arXiv:2209.03914}

\bibitem{Grazzini:2018bsd}
M.~Grazzini, G.~Heinrich, S.~Jones, S.~Kallweit, M.~Kerner, J.M. Lindert,
  J.~Mazzitelli, \emph{{Higgs boson pair production at NNLO with top quark mass
  effects}}.
\newblock \href{https://doi.org/10.1007/JHEP05(2018)059}{JHEP \textbf{05}, 059
  (2018)}, \href{https://arxiv.org/abs/1803.02463}{arXiv:1803.02463}

\bibitem{Baglio:2020wgt}
J.~Baglio, F.~Campanario, S.~Glaus, M.~M\"uhlleitner, J.~Ronca, M.~Spira,
  \emph{{$gg\to HH$ : Combined uncertainties}}.
\newblock \href{https://doi.org/10.1103/PhysRevD.103.056002}{Phys. Rev. D
  \textbf{103}(5), 056002 (2021)},
  \href{https://arxiv.org/abs/2008.11626}{arXiv:2008.11626}

\bibitem{Djouadi:1994ge}
A.~Djouadi, P.~Gambino, \emph{{Leading electroweak correction to Higgs boson
  production at proton colliders}}.
\newblock \href{https://doi.org/10.1103/PhysRevLett.73.2528}{Phys. Rev. Lett.
  \textbf{73}, 2528 (1994)},
  \href{https://arxiv.org/abs/hep-ph/9406432}{arXiv:hep-ph/9406432}

\bibitem{Aglietti:2004nj}
U.~Aglietti, R.~Bonciani, G.~Degrassi, A.~Vicini, \emph{{Two loop light fermion
  contribution to Higgs production and decays}}.
\newblock \href{https://doi.org/10.1016/j.physletb.2004.06.063}{Phys. Lett. B
  \textbf{595}, 432 (2004)},
  \href{https://arxiv.org/abs/hep-ph/0404071}{arXiv:hep-ph/0404071}

\bibitem{Degrassi:2004mx}
G.~Degrassi, F.~Maltoni, \emph{{Two-loop electroweak corrections to Higgs
  production at hadron colliders}}.
\newblock \href{https://doi.org/10.1016/j.physletb.2004.09.008}{Phys. Lett. B
  \textbf{600}, 255 (2004)},
  \href{https://arxiv.org/abs/hep-ph/0407249}{arXiv:hep-ph/0407249}

\bibitem{Actis:2008ts}
S.~Actis, G.~Passarino, C.~Sturm, S.~Uccirati, \emph{{NNLO Computational
  Techniques: The Cases H ---\ensuremath{>} gamma gamma and H ---\ensuremath{>}
  g g}}.
\newblock \href{https://doi.org/10.1016/j.nuclphysb.2008.11.024}{Nucl. Phys. B
  \textbf{811}, 182 (2009)},
  \href{https://arxiv.org/abs/0809.3667}{arXiv:0809.3667}

\bibitem{Actis:2008ug}
S.~Actis, G.~Passarino, C.~Sturm, S.~Uccirati, \emph{{NLO Electroweak
  Corrections to Higgs Boson Production at Hadron Colliders}}.
\newblock \href{https://doi.org/10.1016/j.physletb.2008.10.018}{Phys. Lett. B
  \textbf{670}, 12 (2008)},
  \href{https://arxiv.org/abs/0809.1301}{arXiv:0809.1301}

\bibitem{Muhlleitner:2022ijf}
M.~M\"uhlleitner, J.~Schlenk, M.~Spira, \emph{{Top-Yukawa-induced Corrections
  to Higgs Pair Production}}.
\newblock \href{https://arxiv.org/abs/2207.02524}{arXiv:2207.02524}

\bibitem{Davies:2022ram}
J.~Davies, G.~Mishima, K.~Sch\"onwald, M.~Steinhauser, H.~Zhang, \emph{{Higgs
  boson contribution to the leading two-loop Yukawa corrections to $gg\to
  HH$}}.
\newblock \href{https://arxiv.org/abs/2207.02587}{arXiv:2207.02587}

\bibitem{Dulat:2015mca}
S.~Dulat, T.J. Hou, J.~Gao, M.~Guzzi, J.~Huston, P.~Nadolsky, J.~Pumplin,
  C.~Schmidt, D.~Stump, C.P. Yuan, \emph{{New parton distribution functions
  from a global analysis of quantum chromodynamics}}.
\newblock \href{https://doi.org/10.1103/PhysRevD.93.033006}{Phys. Rev. D
  \textbf{93}(3), 033006 (2016)},
  \href{https://arxiv.org/abs/1506.07443}{arXiv:1506.07443}

\end{thebibliography}

\end{document}